%% file: icml.tex
\documentclass{article}

\usepackage{microtype}
\usepackage{graphicx}
\usepackage{subfigure}
\usepackage{booktabs} %

\usepackage{hyperref}

\usepackage[accepted]{icml2023}

\usepackage{amsmath}
\usepackage{amssymb}
\usepackage{mathtools}
\usepackage{amsthm}

\usepackage[capitalize,noabbrev]{cleveref}

\crefname{definition}{Definition}{Definitions}

\theoremstyle{plain}

\theoremstyle{definition}

\theoremstyle{remark}

\input{resources/math_commands.tex}

\input{include/0-commands.tex}

\usepackage{hyperref}
\usepackage{url}
\usepackage{todonotes}
\usepackage{comment}
\usepackage{booktabs}
\usepackage{mathbbol}
\usepackage{wrapfig}
\usepackage{multirow}
\usepackage{lipsum,booktabs}
\usepackage{xspace}
\usepackage{amsthm}
\usepackage{siunitx}
\usepackage{tabularx}
\usepackage{graphicx}
\usepackage{adjustbox}
\usepackage{amssymb}
\usepackage{enumitem}
\usepackage{pifont}
\newcommand{\cmark}{\ding{51}}%
\newcommand{\xmark}{\ding{55}}%
\DeclareSIUnit\calorine{cal}
\DeclareSIUnit\kcal{\kilo\calorine}
\DeclareSIUnit\angstrom{\text {Å}}

\icmltitlerunning{Uncertainty Estimation for Molecules}

\begin{document}

\twocolumn[
\icmltitle{Uncertainty Estimation for Molecules: Desiderata and Methods}

\icmlsetsymbol{equal}{*}

\begin{icmlauthorlist}
\icmlauthor{Tom Wollschl\"ager}{yyy}
\icmlauthor{Nicholas Gao}{yyy}
\icmlauthor{Bertrand Charpentier}{yyy}
\icmlauthor{Mohamed Amine Ketata}{yyy}
\icmlauthor{Stephan G\"unnemann}{yyy}
\end{icmlauthorlist}

\icmlaffiliation{yyy}{Department of Computer Science \& Munich Data Science Institute, Technical University of Munich, Germany}

\icmlcorrespondingauthor{Tom Wollschl\"ager}{tom.wollschlaeger@tum.de}

\icmlkeywords{Machine Learning, ICML}

\vskip 0.3in
]

\printAffiliationsAndNotice{}  %

\begin{abstract}

Graph Neural Networks (GNNs) are promising surrogates for quantum mechanical calculations as they establish unprecedented low errors on collections of molecular dynamics (MD) trajectories.
Thanks to their fast inference times they promise to accelerate computational chemistry applications.
Unfortunately, despite low in-distribution (ID) errors, such GNNs might be horribly wrong for out-of-distribution (OOD) samples.
Uncertainty estimation (UE) may aid in such situations by communicating the model's certainty about its prediction.
Here, we take a closer look at the problem and identify six key desiderata for UE in molecular force fields, three `physics-informed' and three `application-focused' ones.
To overview the field, we survey existing methods from the field of UE and analyze how they fit to the set desiderata.
By our analysis, we conclude that none of the previous works satisfies all criteria.
To fill this gap, we propose \ours{} (\oursacro{}) a Gaussian Process (GP)-based extension to existing GNNs satisfying the desiderata.
In our extensive experimental evaluation, we test four different UE with three different backbones and two datasets.
In out-of-equilibrium detection, we find \oursacro{} yielding up to 2.5 and 2.1 times lower errors in terms of AUC-ROC score than dropout or evidential regression-based methods while maintaing high predictive performance.
\end{abstract}

\input{include/introduction.tex}
\input{include/desiderata.tex}
\input{include/survey.tex}
\input{include/method.tex}
\input{include/related_work.tex}

\input{include/experiments.tex}
\input{include/conclusion.tex}

\bibliography{bibliography}
\bibliographystyle{icml2023}

\newpage
\appendix
\onecolumn
\input{include/appendix.tex}
\end{document}

%% file: resources/math_commands.tex
\usepackage{amsmath,amsfonts,bm}

\newcommand{\matr}[1]{\bm{#1}}
\newcommand{\vect}[1]{\mathbf{#1}}
\newcommand{\svect}[1]{\bm{#1}}
\newcommand{\set}[1]{\mathcal{#1}}
\newcommand{\normal}[0]{\mathcal{N}}
\newcommand{\gp}[0]{\set{GP}}
\DeclareMathOperator{\EX}{\mathbb{E}}%

\def\eqref#1{equation~\ref{#1}}

\def\1{\bm{1}}

\DeclareMathAlphabet{\mathsfit}{\encodingdefault}{\sfdefault}{m}{sl}
\SetMathAlphabet{\mathsfit}{bold}{\encodingdefault}{\sfdefault}{bx}{n}

\newcommand{\E}{\mathbb{E}}

\newcommand{\KL}{D_{\mathrm{KL}}}

\newcommand{\Cov}{\mathrm{Cov}}

%% file: include/0-commands.tex
\def\pos{\matr{X}}
\def\inputfeatures{\matr{H}}
\def\molecule{\matr M}
\def\dataset{\matr{M}}
\def\energy{E}
\def\forces{\matr{F}}
\def\natoms{n}
\def\featuredim{h}

\def\ndatapoints{N}
\def\ninducingpoints{m}

\def\uncertainty{u}
\def\target{\vect{y}}
\def\pfun{\vect{f}} %
\def\inducingpos{\matr \Phi}
\def\inducingvars{\vect u}

\DeclareMathOperator{\real}{\mathbb{R}}

\DeclareMathOperator{\variance}{Var}
\DeclareMathOperator{\covariance}{Cov}

\DeclareMathOperator{\transpose}{\text{T}}
\DeclareMathOperator{\trace}{\text{Tr}}
\DeclareMathOperator{\groundtruth}{\text{gt}}

\def\ours{Localized Neural Kernel}
\def\oursacro{LNK}
\def\desone{\emph{Symmetry}}
\def\destwo{\emph{Energy conservation}}

\def\desfour{\emph{Locality}}
\def\desfive{\emph{Accuracy}}
\def\dessix{\emph{Confidence-aware}}
\def\desseven{\emph{Speed}}

%% file: include/introduction.tex
\section{Introduction}
In recent years, access to molecular forces has become an essential aspect in various applications such as geometry optimization, and molecular dynamics (MD) simulations~\citep{jensenMolecularModelingBasics2010}.
However, the underlying quantum mechanical (QM) calculations required for these predictions are computationally demanding.
In order to reduce this computational burden, graph neural networks (GNNs) fitted to QM data have been proposed as a means to accelerate MD simulations~\citep{chmiela2017machine,schuttSchNetDeepLearning2018}.
While such surrogates have recently achieved exceptional reproduction of the dataset they have been trained on,  they tend to perform poorly on out-of-distribution (OOD) data \citep{li_out--distribution_2022}. In practical applications, having access to the full-dimensional potential energy surface of molecules is rare. For example, in MD simulations, one typically starts with an initial structure and iteratively applies molecular forces~\citep{hojaQM7XComprehensiveDataset2021}. These simulations may cause the structure to step out of the training domain, rendering the network's predictions unreliable~\citep{stockerHowRobustAre2022}.
Uncertainty estimation (UE) is a promising direction for detecting such unforeseen events.

In the context of molecular force fields, UE comes with a unique and more stringent set of requirements compared to other fields. We categorize these requirements into `physical-informed' and `application-focused' desiderata.
In a survey, we then analyze existing works in UE for molecular force fields on these desiderata.
Our analysis reveals that none of the previous work completely satisfies all desiderata.
To fill the gap, we present \ours{} (\oursacro{}), a Gaussian Process (GP)-based extension to existing GNN-based force fields that reliably estimates uncertainty with a single forward pass while not harming the predictive performance fulfilling all desiderata. 
When testing out-of-equilibrium detection, we find \oursacro{} yielding up to 2.5 and 2.1 times lower errors in terms of AUC-ROC score than dropout-based or evidential-based UE~\citep{gal2016dropout,soleimany2021evidential}.

To summarize, our contributions are:\footnote{Find our code at \href{https://www.cs.cit.tum.de/daml/uncertainty-estimation-for-molecules-desiderata-and-methods}{cs.cit.tum.de/daml/uncertainty-for-molecules}}
\setlist{nolistsep}
\begin{itemize}[noitemsep]
    \item We derive \emph{physics-informed} and \emph{application-focused} desiderata for uncertainty-aware molecular force fields.
    \item We survey previous UE methods based on our desiderata and conclude that existing methods fail on at least one of our desiderata.
    \item We present \ours{} (\oursacro{}), a GP-based extension to existing GNN-based force fields satisfying all desiderata.
\end{itemize}

%% file: include/desiderata.tex
\section{Uncertainty Estimation Criteria for Molecular Predictions}
\label{sec:approach}

We consider the task of energy and force predictions on molecules. One molecule is represented as a point cloud of $\natoms$ points (atoms), each associated with a position and a set of rotationally invariant features (e.g., atom types), defined as $\matr X \in \real^{\natoms \times 3}$ and $\matr H \in \real^{\natoms \times \featuredim}$, respectively.
In addition to learning the molecular energy $\energy \in \real$ and the atom forces $\forces \in \real^{\natoms \times 3}$, we investigate the ability of our approach, \oursacro{},  to quantify energy uncertainty $\uncertainty_{\energy}$ and demonstrate that it can further be used in a variation-based way to quantify force uncertainty $\uncertainty_{\forces}$. 
In particular, we aim at learning a function $f_\theta$ such that $f_\theta(\pos, \inputfeatures)\approx\energy^{\groundtruth}$ and $\frac{\partial f_\theta}{\partial \matr X}(\pos, \inputfeatures) \approx \forces^{\groundtruth}$ where  $\energy^{\groundtruth}$ and $\forces^{\groundtruth}$ denote the ground-truth molecular energy and forces, respectively. The predicted values are denoted with star subscript, i.e., $\energy_*$ and $\forces_*$. We denote the training set consisting of all molecule positions and features as $\set\dataset = \{(\pos_1, \inputfeatures_1), \dots, (\pos_\ndatapoints, \inputfeatures_\ndatapoints)\}$.
Stacking the tuples of the dataset into a matrix, we refer to it as $\dataset$ with corresponding target vector of energies $\target$.

\subsection{Desiderata}\label{sec:desiderata}
Dealing with energy and forces in molecular structures, which are influenced by physical symmetries, poses unique challenges for UE in the context of molecular force fields. To effectively perform UE in this area, it is essential to take into account both \emph{physics-informed} and \emph{application-focused} desiderata. Physics-informed desiderata capture the physical constraints and behavior characteristic of molecular systems. In this work, we identify three main physics-informed desiderata:
\begin{itemize}
    \item \desone{}: Symmetries play an essential role in physics. They describe the fundamental behavior of quantities such as energies or forces under euclidean transformations. For instance, the energy of a molecule $\energy$ is invariant to the euclidean group $\delta\in\mathcal{E}(3)$ and permutation group $\pi\in S_n$, i.e., $\energy(\delta\circ\pi\circ\pos,\pi\circ\inputfeatures)=\energy(\pos,\inputfeatures)$, while the associated atomic forces behave equivariantly, $\forces(\delta\circ\pi\circ\pos,\pi\circ\inputfeatures)=\delta\circ\pi\circ\forces(\pos,\inputfeatures)$. 
    \item \destwo{}: Molecular forces, being the gradient of an energy surface, must form a conservative vector field, i.e., all paths from point a to b have the same integral and the force field must be curl-free.
    One may obtain all these properties by defining the forces via differentiation of a scalar field $\forces:=\frac{\partial\energy}{\partial\pos}$.
    \item \desfour{}: As most of the interaction between atoms happens within a short distance, locality plays an important role in molecular force fields~\citep{leachMolecularModellingPrinciples2001}. While there are long-range interactions in molecules, these only account for a small fraction of the total energy and are negligible for molecular forces~\citep{lanAdsorbMLAcceleratingAdsorption2023}.
    As a consequence of locality, a molecule's energy behaves size-consistently with increased system sizes.
    Formally, $\energy \approx \sum_{i=1}^n \energy_i$, where $\energy$ is the total energy of $n$ molecules and $\energy_i$ is the energy of the $i$th molecule. Similarly, the energy's uncertainty should behave additive under extension.
\end{itemize}
With MD simulations as an application in mind, we identify three key application-focused desiderata:
\begin{itemize}
    \item \desfive{}: MD simulations require high precision and are often performed for millions of steps where each step simulates femto seconds ($10^{-15}\si{\second}$)~\citep{stockerHowRobustAre2022}. To keep simulations stable and not accumulate errors, high predictive performance and, thus, accurate reproduction of the QM calculations is essential.
    \item \dessix{}: During an MD simulation, a structure may move outside of the training domain. In such cases, the surrogate must effectively communicate its uncertainty regarding its predictions. If this aspect is not accounted for, divergence has been observed~\citep{stockerHowRobustAre2022}.
    \item \desseven{}: Lastly, as MD simulations typically involve millions of steps \cite{Narumi1999}, it is important to preserve the efficient runtime of surrogate methods.
\end{itemize}

As we discuss in \autoref{sec:survey}, current UE approaches for molecules lack at least one of these desiderata. We address this by introducing \oursacro{}, a GP-based extension to existing GNNs that satisfies all desiderata.

%% file: include/survey.tex
\section{Survey of UE for molecular force fields}\label{sec:survey}
\begin{table*}
    \centering
    \resizebox{\linewidth}{!}{
    \input{tables/desiderata_fulfillment2.tex}
    }
    \caption{Desiderata fulfillment of different uncertainty estimation methods.}
    \label{tab:comp_methods}
\end{table*}
While there is a variety of UE methods for deep learning\footnote{we refer the interested reader to \citet{gawlikowski2021survey}}, here we only discuss (twice) differentiable solutions that fit the application of predicting molecular energies and forces.
UE methods can be broadly categorized into two families of methods: \emph{sampling-based} and \emph{sampling-free} methods.
For each method, we outline its advantages and disadvantages and analyze it based on our desiderata set in \autoref{sec:desiderata}.
An overview of the methods can be found in \autoref{tab:comp_methods}.

\textbf{Ensembles} are touted as the `gold-standard' solution when it comes to UE.
An ensemble is a collection of different methods trained on the same dataset \cite{dataset-shift}.
By varying between architecture or initialization between the individual models, one obtains several estimates about the desired property.
By computing statistical metrics of these samples, e.g., their standard deviation, one obtains an uncertainty quantification \cite{ensembles}. In mathematical notation this can be expressed as: 
\begin{equation}\label{eq:ensemble_uncertainty_estimate}\begin{aligned}
    &\uncertainty_{\energy} = \sqrt{\variance \left(f_1(\molecule), \dots, f_N(\molecule)\right)}\\
    &\uncertainty_{\forces} = \trace \left[\covariance\left(\frac{\partial f_1(\molecule)}{\partial \pos}, \dots \frac{\partial f_N(\molecule)}{\partial \pos} \right) \right].
\end{aligned}\end{equation}
Here $u_{\energy}, u_{\forces}$ are the measures of the predictive uncertainty for the energy and forces. $f_i$ denotes the i-th component of the ensemble, i.e., the i-th trained network. We use the trace as a metric to capture the uncertainty of the forces covariance matrix in a single number.
In terms of our desiderata, ensembles generally preserve the properties of the underlying methods.
Thus, they satisfy the same desiderata as their components with \desseven{} being a notable exception.
As one must train several models \emph{and} perform inference multiple times, an ensemble's runtime scales linearly with its size. 
In case one uses ensembles of GNNs~\citep{gasteigerDirectionalMessagePassing2019} one directly satisfies the physics-informed. 
As we show in \autoref{sec:experiments}, ensembles also empirically perform well on \desfive{} and \dessix{} but crucially fail at \desseven{}.

\textbf{Monte Carlo Dropout} (MCD) is proposed as a comparatively efficient alternative to ensembles with the idea of merging the ensemble into a single model with probabilistic output. Instead of having different models as in ensembles, we leverage the dropout during inference and get multiple predictions for the same molecule. Hence, the formulation for $u_{\energy}, u_{\forces}$ is the same as in \autoref{eq:ensemble_uncertainty_estimate} while the only difference is that $f_i$ denotes a single evaluation of the same model but with different dropped connections \cite{gal2016dropout}. As one simply drops neurons within one model, only one model has to be trained.
Like ensembles, MCD generally inherits the desiderata from its base model but only in expectation as it randomly drops connections.
As we show in \autoref{sec:experiments}, MCD deteriorates the predictive performance.
Further, we observe that while in expectation locality preserving, MCD's uncertainty is not behaving size-consistent for practical sample sizes and is thus not fulfilling \desfour{}.
Lastly, while one only has to train a single model compared to ensembles, one still must perform multiple passes during inference violating the \desseven{} desiderata.

\textbf{Evidential Regression}:
In evidential regression, one directly parametrizes an evidential distribution over the target rather than the target~\citep{soleimany2021evidential}.
In practice, this means that instead of directly estimating a normal distribution on the energy, the model estimates the parameters of an evidential Normal-Inverse-Gamma distribution \cite{evidential-regression,charpentier_natural_2022}. 
Given this distribution, one can perform statistical tests to quantify uncertainties.
While only requiring a single forward pass, this approach does not necessarily yield trustworthy uncertainty estimates, as we will see in \autoref{sec:experiments}.
Similar to the unreliability of the energy prediction itself, the outputted distribution parameters are similarly subject to unpredictable behavior outside the training regime \cite{charpentierPosteriorNetworkUncertainty2020, dbu-robustness}.
Further, by predicting a single distribution for the entire molecular graph, evidential regression requires global embeddings violating \desfour{}.

\begin{figure*}[t!]
    \centering
    \includegraphics[width=\textwidth]{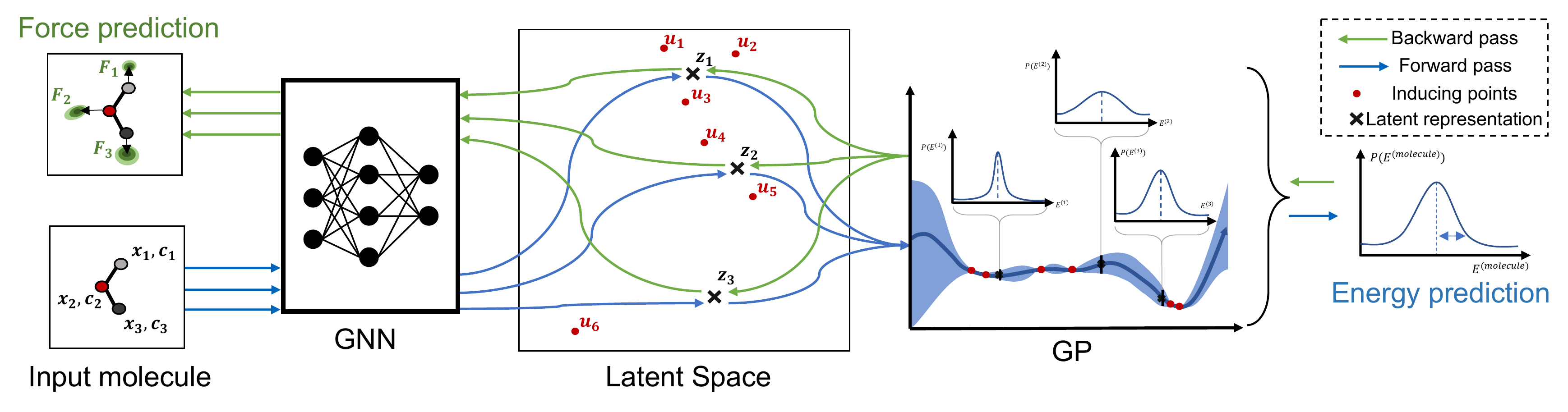}
    \vspace{-0.15in}
    \caption{Overview of \ours{}}
    \label{fig:model_figure}
    \vspace{-0.15in}
\end{figure*}

\textbf{Gaussian Processes} (GPs) are a type of non-parametric models for distributions over functions that are known for their flexibility and expressiveness \cite{bishop, Jakkala2021DeepGP}. 
They are particularly frequently used in UE.
One of the key advantages of using GPs is that they allow for the modeling of complex relationships between variables, while also providing a way to quantify the uncertainty associated with the predictions made by the model.
In the following, we adapt the notation from \citet{Rasmussen2009GaussianPF} to our setting.
Given the data $\dataset$, we assume a GP prior: $f(\dataset) \sim \gp(\mu(\dataset), k_\gamma(\dataset, \dataset^{\transpose}))$, where $\mu(\cdot)$ is the mean function and $k_\gamma(\cdot, \cdot)$ is the covariance function with parameters $\gamma$. Any collection of points $\pfun$ lying on $f(\dataset)$ follows a normal distribution as:
\begin{equation}
\begin{aligned}
	p(\pfun) &= p(f(\dataset)) \\&= p(\left[f(\dataset_1), \dots, f(\dataset_n) \right]^\top)  \sim \normal (\pfun \vert\, \svect \mu, \matr K_{\dataset \dataset}),
 \end{aligned}
\end{equation}
where $\svect \mu_i = \mu(\dataset_i)$ is the mean vector and $\matr K^{\dataset, \dataset}_{ij} = k_\gamma(\dataset_i, \dataset_j)$ the Gram matrix. Assuming that our labels are given as noisy observations of the latent variable, we model the prediction as: $\vect y_i = f(\dataset_i) + \epsilon$, where $\epsilon \sim \normal(0, \sigma)$. We can then predict the values for a new collection of inputs, $\dataset_\star$, via the predictive distribution:
\begin{equation}
	p(\pfun_\star \,\vert \, \dataset_\star, \dataset, \vect y, \sigma, \gamma) \sim \normal (\pfun_\star \vert\, \EX [\pfun_\star], \Cov(\pfun_\star))
\end{equation}
\begin{equation}
	\EX [\vect {f}_\star] = \mu_{\dataset_\star} + \matr K_{\dataset_\star \dataset} \left[K_{\dataset \dataset} + \sigma ^2 \matr I \right]^{-1} \vect y, \quad 
\end{equation}
\begin{equation}
	\Cov(\pfun_\star) =  \matr K_{\dataset_\star \dataset_\star} - \matr K_{\dataset_\star \dataset}\left[\matr K_{\dataset \dataset} + \sigma ^2 \matr I\right] ^{-1}\matr K_{\dataset \dataset_\star}.
\end{equation}
As shown by \citet{vandermause2020On}, we can define a GP acting on the input and hand-crafted features to predict the forces of a molecule with an energy-conserving model while being rotationally covariant. However, due to the inversion of the Gram matrix, the exact GP inference scales with $\mathcal{O}(\ndatapoints^3)$ where $\ndatapoints$ is the number of columns or rows of $\dataset$. Furthermore, compared to recently proposed GNNs the GPs only achieve low accuracies~\citep{gasteiger2021gemnet,batznerEquivariantGraphNeural2022}. 

The desiderata \desfive{} and \desseven{} are thus not fulfilled. 
Further, because a GP compares the graphs as a whole, it does not scale additively with system size, thereby violating \desfour{}.

\textbf{Stochastic Variational Gaussian Process} (SVGP): As the exact GP requires the inversion of the covariance matrix, it comes with a significant computational burden. To mitigate this issue, sparse approximations and variational inference have been proposed, known as Sparse and Variational GP (SVGP) \citep{titsias2009Variational, hensman2013Gaussian, hensman2014Scalable}. 

The SVGP approach entails the selection of a subset of data points, referred to as inducing points $\inducingpos$ and variables $\inducingvars$ to approximate the underlying distribution through the use of variational inference. A prior can be imposed on the joint distribution:\footnote{We adapt \citet{hensman2014Scalable}'s notation for consistency.}
\begin{equation}
	p(\pfun, \inducingvars) \sim \normal \left(
	\begin{aligned}
		\left[
			{\begin{array}{c}
				\pfun\\
				\inducingvars\\
			 \end{array}}
		\right] 
	\end{aligned}\vert\; \vect 0, 
	\begin{aligned}
		\left[
			{\begin{array}{cc}
				\matr K_{\dataset \dataset} & \matr K_{\inducingpos\dataset}\\
				\matr K_{\inducingpos \dataset} & \matr K_{\inducingpos \inducingpos}\\
			\end{array}}
		\right]	
	\end{aligned}
	\right).
\end{equation} 
Note, that this equation describing the relationship between latent variables $\pfun$ and inducing variables $\inducingvars$ is equivalent to the relationship between the training data and the new test points for an exact GP. 

To learn the inducing points and variables, we need to include these into the optimization formulation. \citet{titsias2009Variational} introduced variational inference to approximate the posterior $p(f, u \vert \vect y)$ by $q(\pfun, \inducingvars)$. With the introduction of additional variational parameters, \citet{hensman2013Gaussian} extended this approach and enabled stochastic gradient optimization. Assuming $q(\pfun, \inducingvars) = p(\pfun \vert \inducingvars) q(\inducingvars)$ and $q(\inducingvars)$ being gaussian, we can lower bound the marginal likelihood by the Evidence Lower Bound (ELBO):
\begin{equation}\begin{aligned}\label{eq:elbo}
	\log &p(\pfun_\star) \\ &\geq \EX_{q(\inducingvars)p(\pfun\vert \inducingvars)} \left[\log p(\pfun_\star \vert \pfun)\right] - \KL\left(q(\inducingvars) \vert\vert p(\inducingvars)\right).
\end{aligned}\end{equation}
This ELBO can be used as loss function for stochastic gradient optimization. The computational complexity of an SVGP is in $\mathcal{O}(\ninducingpoints^2\ndatapoints)$ where $m$ is the number of inducing points. Important to note, is that the inducing points and variables are optimized to best fit the distribution and are not instances of the training data. 

Compared to the exact GP, the SVGP fulfills the \desseven{} desideratum. However, as the inducing points are free variables to be optimized, their location information has no connection to the atom positions breaking rotational invariance~\citep{vandermause2020On}).
Thus, compared to GPs, \desone{} is not fulfilled anymore. 

%% file: tables/desiderata_fulfillment2.tex
\definecolor{mygreen}{HTML}{3C8031}
\definecolor{myred}{HTML}{A9341F}

\newcommand{\tick}{{\color{mygreen}\cmark}}
\newcommand{\cross}{{\color{red}\xmark}}
\newcommand{\neutral}{{\color{orange}$\bm{\sim}$}}

\begin{tabular}{lcccccccc}
\toprule
{} & Ensemble & MC-Dropout & Evidential & GP & SVGP & SVGP-DKL & \oursacro{}\\
\midrule
\desone{} & \tick & \tick & \tick & \tick & \cross & \tick & \tick \\
\destwo{} & \tick & \tick & \tick & \tick & \cross & \tick & \tick \\
\desfour{} &  \tick & \neutral & \cross & \cross & \cross & \cross & \tick  \\
\desfive{} & \tick & \neutral & \tick & \cross & \cross & \neutral & \tick \\
\dessix{} & \tick & \tick & \neutral & \tick & \tick & \tick & \tick \\
\desseven{} & \cross & \cross & \tick & \cross & \tick & \tick & \tick \\
\bottomrule
\end{tabular}

%% file: include/method.tex
\section{\ours{}}\label{sec:method}
In the following, we introduce \oursacro{} in two steps.
First, we introduce a non-localized version, termed SVGP-DKL, by combining the SVGP with Deep Kernel Learning (DKL)~\citep{wilson2015Deep}.
In DKL one uses deep neural networks to encode the data in a learned, potentially lower-dimensional, space before applying the GP. Both are then trained end-to-end.
Second, we discuss how we localize the SVGP-DKL to obtain \oursacro{} fulfilling all desiderata.

\textbf{SVGP-DKL}:
To leverage the predictive performance of GNN-based molecular force fields, we use such a GNN as encoder for DKL.
Further, we avoid the expensive runtimes of the GP by relying on the SVGP approximation.
Let $h(\pos, \inputfeatures) = (h_\text{pred} \circ h_{\text{rep}})(\pos, \inputfeatures) $ be a GNN with $h_\text{rep}:\real^{n\times 3} \times \real^{n \times h} \rightarrow \real^{n \times d}$ being its learned invariant atom representations and $h_\text{pred}:\real^{n \times d} \rightarrow \real$ its property predictor.
By replacing the predictor $h_\text{pred}$ with an SVGP, we combine the benefits of GNNs with approximate GPs.
Note that GPs operate on fixed-dimensional spaces and we, thus, have to introduce a summation over all atoms to obtain a representation for the whole molecule:
\begin{equation}
\begin{aligned}
p(E_\star \vert \pos, \inputfeatures) &= \gp_\phi \circ \sum_{i=1}^n h_{\text{rep}}(\pos, \inputfeatures)_i\\ &\sim \normal\left(E_\star \vert \EX [\energy_\star], \Cov(\energy_\star)\right)\label{eq:global_gp}
\end{aligned}
\end{equation}
We can view this as a learned kernel \cite{wilson2015Deep}: 
\begin{equation}
\begin{aligned}
    &k_{\text{SVGP-DKL}}((\matr X^{(k)}, \matr H^{(k)}),(\matr X^{(j)}, \matr H^{(j)})) =\\ k&\left(\sum_i h_{\text{rep}}(\matr X^{(k)}, \matr H^{(k)})_i, \sum_i h_{\text{rep}}(\matr X^{(j)}, \matr H^{(j)})_i\right)
\end{aligned}
\end{equation}
Hence, the prediction of the model is a normal distribution. We can use the variance of the predictive distribution as energy uncertainty estimate. Given that the embeddings of the studied GNNs (SchNet, DimeNet++, NequIP) are invariant to the euclidean group $\set E(3)$ and permutations $S_n$, this GP extension fulfills \desone{}.  

To fulfill the physical desideratum of \destwo{}, we need to calculate the forces as the derivative of the energy. Hence, our method needs to be twice differentiable to be trained. Taking the derivative of the energy with respect to the position thus yields the force prediction: 
\begin{equation}
\matr F = -\frac{\partial E}{\partial \pos} = -\frac{\partial \text{GP}_\phi}{\partial h_{\text{rep}}}\frac{\partial h_{\text{rep}}}{\partial \pos}. 
\end{equation}
As a GP with RBF kernel is $C^\infty$ smooth, we only have to pick a twice differentiable representation function $h_{\text{rep}}$ to fulfill \destwo{}. 
To train the GP, we use the predictive log-likelihood as objective to estimate the energy error \cite{jankowiak_parametric_2020}. This objective, similar to the variational ELBO shown in \autoref{eq:elbo}, can be formally written as: 
\begin{equation}
    \mathcal{L} = \log \E_{q(\vect u)p(\pfun|\vect u)}[p(\pfun_*|\pfun)] - D_{\text{KL}(q(\vect u)||p(\vect u))}.
\end{equation}
This formulation has been reported to yield improved estimates of predictive variances when compared to alternative approaches \cite{jankowiak_parametric_2020, 9982167}. We use the RMSE loss to estimate the force error.
However, as the GP operates on the singular, fixed-size, molecule embedding, it fails to satisfy \desfour{}.

\textbf{Localizing SVGP-DKL}:
To attain \desfour{}, we use a localized version of the just defined GNN-based GP to obtain \ours{} (\oursacro{}). 
We achieve this by reversing the order of the sum and the GP in \autoref{eq:global_gp}.
Instead of using the global graph embedding, we fit the GP on the atomic embeddings directly and our final output is the sum of these atomic GPs. Hence, the predictive distribution of the predicted energy value $\energy_*$ can be written as\footnote{Assuming independence of the atom random variables and, thus, summing only the diagonal elements leads to better performance.}: 
\begin{align}
    p(E_\star \vert \pos, \inputfeatures) &= \sum_i^\natoms \gp_\phi \circ h_{\text{rep}}(\pos, \inputfeatures)_{i}\\  
    &\sim \normal\left(E_\star \,\vert\,\sum_i^\natoms \EX [\energy_\star]_i, \sum_{ij}^\natoms \Cov(\energy_\star)_{ij}\right).\nonumber
\end{align}
Since we sum the individual atom energy contributions and uncertainties, we fulfill \desfour{}. Compared to the global embedding version, adding an independent molecule will increase the uncertainty proportionally. We will show this comparison also empirically in \autoref{sec:experiments}.
In regards to the `application-focused' desiderata, \oursacro{} only needs a single forward pass fulfilling \desseven{}, and as we will show in \autoref{sec:experiments} also fulfills \desfive{} and \dessix{}. 
Thus, it is the only approach fulfilling all of our set desiderata.

\textbf{Avoiding pitfalls.}
As the use of large encoder networks can lead to unstable uncertainty estimation for deep kernel learning models~\citep{ober_promises_2021}, proper regularization has to be deployed.
Ober et al. suggest using a full-bayesian treatment to avoid this pitfall. We use two different approaches: (1) approximate a full bayesian treatment by using dropout \cite{gal2016dropout, srivastava2014dropout} and (2) a fixed-encoder which is motivated by the strong empirical performance in their work (see Table 3 \citet{ober_promises_2021}).

In order to apply dropout to equivariant networks like NequIP, one cannot simply drop any features.
In \autoref{app:baselinedetails}, we discuss how we translate dropout to such group-equivariant neural networks.
Note that during inference, we do not perform dropout with \oursacro{} as we use the variance of the GP's predictive distribution as an uncertainty estimate rather than an empirical Monte Carlo estimate.

Instead of applying dropout, one might also avoid overfitting by fixing the representation function $h_\text{rep}$.
This could be done, e.g., by first training the GNN with its original predictor on the objective.
After pretraining, the representation function will be frozen similar to common fine-tuning on computer vision models.
By limiting the embedding space in such a way we can avoid DKL pitfalls~\citep{ober_promises_2021}.

In the following, we refer to the dropout training approach as \oursacro{} while we call the fixed encoder approach fixed-\oursacro{}.

%% file: include/related_work.tex
\section{Related work}\label{sec:related}

\textbf{Machine learning potentials}:
Machine learning has a long history in learning potentials and force fields for molecular simulations. Starting with \citet{halgrenMerckMolecularForce1996} where empirical force fields were introduced with the goal of fitting functions to QM calculations to avoid performing the expensive QM calculations many times.
Later, kernel methods~\citep{behlerAtomcenteredSymmetryFunctions2011,bartokRepresentingChemicalEnvironments2013,christensenFCHLRevisitedFaster2020} based on sophisticated feature constructions of molecular neighborhoods took over the field.
These kernel methods were fast to evaluate and capable of accurately capturing molecular interactions.
By focusing on locality and freely learnable featurizations GNNs~\citep{schuttSchNetDeepLearning2018,unkePhysNetNeuralNetwork2019} achieved accurate reproductions of molecular force fields.
Due to the implementation of physical symmetries to the euclidean group $\set E(3)$ and the permutational group $S_n$, GNNs presented sample efficient ways of learning from molecular data.
Recent advances provided universal models, i.e., models that can model any function on geometry point clouds, accomplished by including dihedral angles~\citep{gasteigerGemNetUniversalDirectional2021,liuSphericalMessagePassing2021} or by the introduction of SO(3) equivariant models~\citep{thomasTensorFieldNetworks2018,batznerEquivariantGraphNeural2022}.
Other developments focus on closing the gap between QM calculations and ML potentials, e.g., by including prior physical knowledge~\citep{unkeSpookyNetLearningForce2021}, by providing efficient QM calculations as additional input~\citep{qiaoOrbNetDeepLearning2020,qiaoUNiTEUnitaryNbody2021}, by implementing DFT functionals with GNNs~\citep{snyderFindingDensityFunctionals2012,kirkpatrickPushingFrontiersDensity2021}, or by learning potentials directly from first principle~\citep{gaoAbInitioPotentialEnergy2022,gaoGeneralizing2023,gaoSamplingfreeInferenceAbInitio2022}.

\textbf{Uncertainty in machine learning.} There are different approaches for uncertainty estimation (UE) in deep learning. The interested reader is referred to \citet{gawlikowski2021survey} for a detailed survey. UE methods can be broadly categorized into two families of methods: \emph{sampling-based} and \emph{sampling-free} methods. 

Sampling-based methods generally evaluate the uncertainty by estimating statistics over multiple different predictions. Generally, when the prediction is very similar of over the different samples, this indicates the low-uncertainty of the model. Ensembles \citep{bayesian-classifier-combination,ensembles, dynamic-bayesian-combination-classifiers,batch-ensembles,hyper-ensembles}, often refered to as the gold-standard, are a collection of individual models. They estimate the uncertainty by the variation in the prediction of the different ensemble members. Dropout methods \cite{gal2016dropout}, approximate this technique by following the argument that dropout let the model learn different subnetworks during training. Thus each random initialization might be an approximation of an ensemble member. Other notable techniques are bayesian neural networks \citep{bayesian-networks, scalable-laplace-bnn, simple-baseline-uncertainty}, where the model parameters are a distribution instead of a single value. Hence, one can sample parameters of the model multiple times and use these to infer a data sample resulting in multiple predictions from which we can calculate the variance again. 

Sampling-free generally requires to evaluate the uncertainty in a single forward-pass methods. Evidential methods propose to parametrize conjugate prior distributions \citep{survey_evidential_uncertainty,dbu-robustness,max_gap_id_ood,uncertainty-generative-classifier,multifaceted_uncertainty,graph_posterior, lightweight-prob-net}. Deep Kernel Learning aim at learning Gaussian processes in latent space with random feature projections or learned inducing points \citep{uncertainty-distance-awareness, due, duq, uceloss}. Calibration models are methods which aim at predicting confidence which are good approximations of the true probability of correctness \citep{accurate-uncertainties-deep-learning-regression, confidence-aware-learning, individual-calibration, distribution-calibration-regression, intra-order-preserving}. Other models propose to propagate uncertainty accross layers and model uncertainty at the weight or activation levels \citep{natural-parameter-network, sampling-free-variance-propagation, feed-forward-propagation, lightweight-prob-net, probabilistic-backprop-scalable-bnn}. Most of these models do not predict uncertainty on molecular properties. Only \citet{soleimany2021evidential} leverage uncertainty estimation of evidential deep learning methods for guided molecular property prediction and show its effectiveness. We use their work as a further baseline, to which we compare our results. Another interesting approach is the Graph Mixture Density Network \cite{errica2021graph}, which outputs a mixture of distributions on the predicted target but does not focus on molecules.

%% file: include/experiments.tex
\section{Experiments}
\label{sec:experiments}
In this section, we focus on the empirical evaluation of our desiderata.
We compare Ensembles, Monte Carlo dropout (MCD), evidential regression, SVGP-DKL\footnote{For SVGP-DKL versus LNK, see \autoref{app:svgp_vs_lnk}}, and \oursacro{}.
We pair each of the UE methods with SchNet~\citep{schutt2021equivariant}, DimeNet++\citep{gasteigerDirectionalMessagePassing2019,gasteigerFastUncertaintyAwareDirectional2020} and NequIP~\citep{batznerEquivariantGraphNeural2022}.
As traditional GPs and SVGP either do not allow for varying input sizes or result in poor predictive performance, we do not discuss these here.
Further, as \desone{} and \destwo{} are fixed desiderata, discussed in earlier \autoref{sec:survey} and \autoref{sec:method}, we do not perform experiments for those.

In our first experiment, we analyze \desfour{} by testing whether the UE is size consistent.
Next, we analyze the \desfive{} on two common QM datasets, MD17~\citep{chmiela2017machine} and QM7X~\citep{hojaQM7XComprehensiveDataset2021}.
To measure \dessix{}, we look at the calibration score and out-of-distribution (OOD) detection.
For calibration, we use the calibration regression score from \citet{charpentier_natural_2022}.
For OOD detection we use the area under the receiver operating characteristic curve (AUC-ROC). For each backbone, we use the respective hyperparameters from their papers, listed in \autoref{app:baselinedetails}.
If not further specified, we train each model on the standard combined energy and force loss~\citep{gasteigerDirectionalMessagePassing2019}: 
\begin{equation}
\mathcal{L}_\text{total} = (1-\rho)\mathcal{L}_\text{energy} + \rho\mathcal{L}_\text{force}.
\end{equation}
Here $\rho$ is a hyperparameter to set the importance of energy and force loss. For $\text{\energy}, \text{force}$, we specify the loss in \cref{app:baselinedetails}.

\textbf{Methods.}
\underline{\emph{Ensembles:}} For Ensembles we use a sample size of $5$ differently initialized models trained on the same dataset.
\underline{\emph{Monte-Carlo Dropout:}} Similar to Ensembles, we choose a sample size of 5 for evaluating uncertainties. 
As dropout rate, we use a rate of \SI{20}{\percent}.
We study different dropout rates in \autoref{app:dropout_exp} but found \SI{20}{\percent} to be good tradeoff between \desfive{} amd \dessix{}.
\underline{\emph{Evidential regression:}} We adapt the energy loss according to \cite{soleimany2021evidential} instead of the MAE. Note that we use a smaller coefficient for the forces loss of $\rho=0.9$ (instead of $\rho=0.99$ used with other models).
In early experiments, we found larger force coefficients leading to poor uncertainty estimates.
\underline{\emph{SVGP-DKL and \oursacro{}:}} As described in \autoref{sec:method}, we use the Predictive Log-Likelihood as energy loss instead of the MAE. We compare \oursacro{} against SVGP-DKL in \cref{app:svgp_vs_lnk} to highlight the importance of localization. \underline{Graph Mixture Density Networks:} we investigate the performance of this approach in \cref{app:gmdn}.

\textbf{Datasets.}
\underline{\emph{QM7-X:}} \cite{hojaQM7XComprehensiveDataset2021} This dataset covers both equilibrium and non-equilibrium structures. We train on equilibrium structures and non-equilibrium structures are considered OOD data. \underline{\emph{MD17:}} \cite{chmiela2017machine} MD17 contains energies and forces for molecular dynamics trajectories of different organic molecules. %

\subsection{\desfour{}}\label{sec:exp_extensivity}
\begin{figure}
    \centering
    \includegraphics[width=\linewidth]{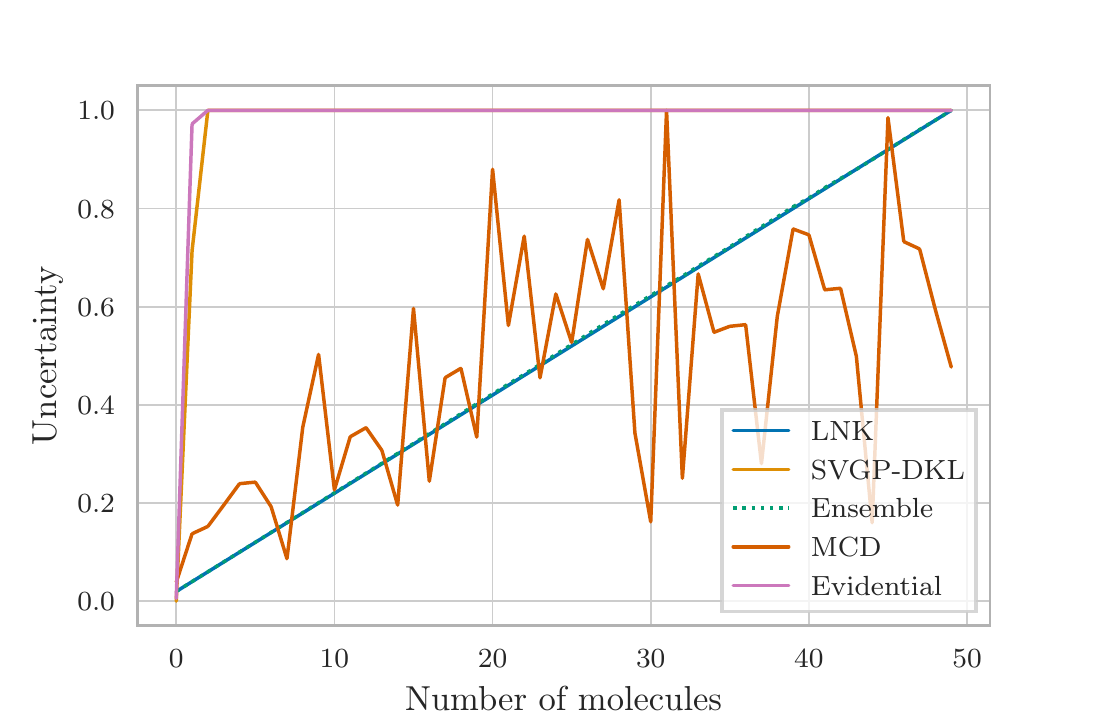}
    \caption{$u_{\energy}$ for increasing number of identical molecules with sufficient distance to one another.}
    \label{fig:extensivity}
\end{figure}
In this experiment, we analyze the \desfour{} property for all UE models.
For this, we duplicate the same molecule translated by \SI{15}{\angstrom}\footnote{The distance larger than the cutoff of all GNNs.}.
We measure the total uncertainty of the increasing number of molecules.
For a model fulfilling \desfour{}, the total uncertainty should be proportional to the number of molecules, i.e., $\uncertainty_\energy \propto i$ where $i$ is the number of copies.

\autoref{fig:extensivity} depicts the dependency of the uncertainty by the number of molecules.
We observe that both SVGP-DKL and Evidential do not exhibit the desired behavior. The increase of uncertainty is too steep and after three molecules, stops increasing. Ensembles and \oursacro{} meet the requirements of increasing linearly with slope $1$. We can observe that MCD exhibits an increase in uncertainty with the growing number of molecules but due to its randomness, \desfour{} will only be satisfied in expectation. 

\subsection{\desfive{}}
\begin{table}
\caption{MAE of SchNet on MD17 with different UE methods (energies in $\si{\kcal\per\mole}$, forces in $\si{\kcal\per\mole\per\angstrom}$)}\label{tab:md17_schnet}
\resizebox{\columnwidth}{!}{
\input{tables/md17/schnet.tex}
}
\end{table}
For this task, we use the fixed-\oursacro{} as it enables us to reuse one ensemble backbone while showing the comparison of how well the pure GNN can perform on that dataset. Hence, we also list the backbone performance as a reference. 
We compare the average predictive performance across all seven MD17 molecules for each combination of backbone and UE method in \autoref{tab:md17_average}.
One may see that while MCD and evidential regression sacrifice accuracy, \oursacro{} maintains backbone-like predictive performance. 
For SchNet, we provide per-molecule results in \autoref{tab:md17_schnet}. We show additional results for other backbones in \autoref{app:additional_experiments}.

\begin{table}
    \centering
    \caption{Average metrics over seven molecules of MD17 for different backbones}
    \vspace{0.05in}
    \label{tab:md17_average}
    \resizebox{\columnwidth}{!}{
    \input{tables/md17/average_comparison.tex}
    }
\end{table}

In \autoref{tab:qm7x_dimenet_predictive}, we can see the predictive performance on QM7X.
In energy prediction, we again observe that Ensembles, Evidential, and \oursacro{} perform well in terms of accuracy.
Remarkably, MCD's energy accuracy significantly deteriorates with increasing dropout rates.
For forces, Ensembles, and \oursacro{} perform well. But, unlike energies, the Evidential approach significantly worsens predictive results.
Despite its worse energy error, MCD achieves low force errors.

\begin{table}
    \caption{MAE of Dimenet++ trained on QM7X-Equilibrium with different UE methods (energies in $\si{\eV}$, forces in $\si{\eV\per\angstrom}$)}
    \vspace{0.05in}
    \resizebox{\columnwidth}{!}{
    \input{tables/qm7x/qm7x_dimenet_predictive.tex}

    }
    \label{tab:qm7x_dimenet_predictive}
\end{table}

\subsection{\dessix{}}

Now, we measure the uncertainty estimates for different approaches. For SchNet on MD17, \autoref{tab:md17_schnet} lists the calibration scores for each UE method on each molecule.
Here, most approaches perform similarly while MCD surprisingly yields the the best results.
We observe a similar pattern for QM7X in \autoref{tab:qm7x_dimenet_predictive}. However, the energy prediction of MCD with $0.24$ is more than a magnitude worse than others and as a uniform distribution minimizes the calibration score, it only has significance if the energy prediction is good. All other methods have similar calibration scores.

\begin{figure}
\centering
\includegraphics[width=1.0\columnwidth]{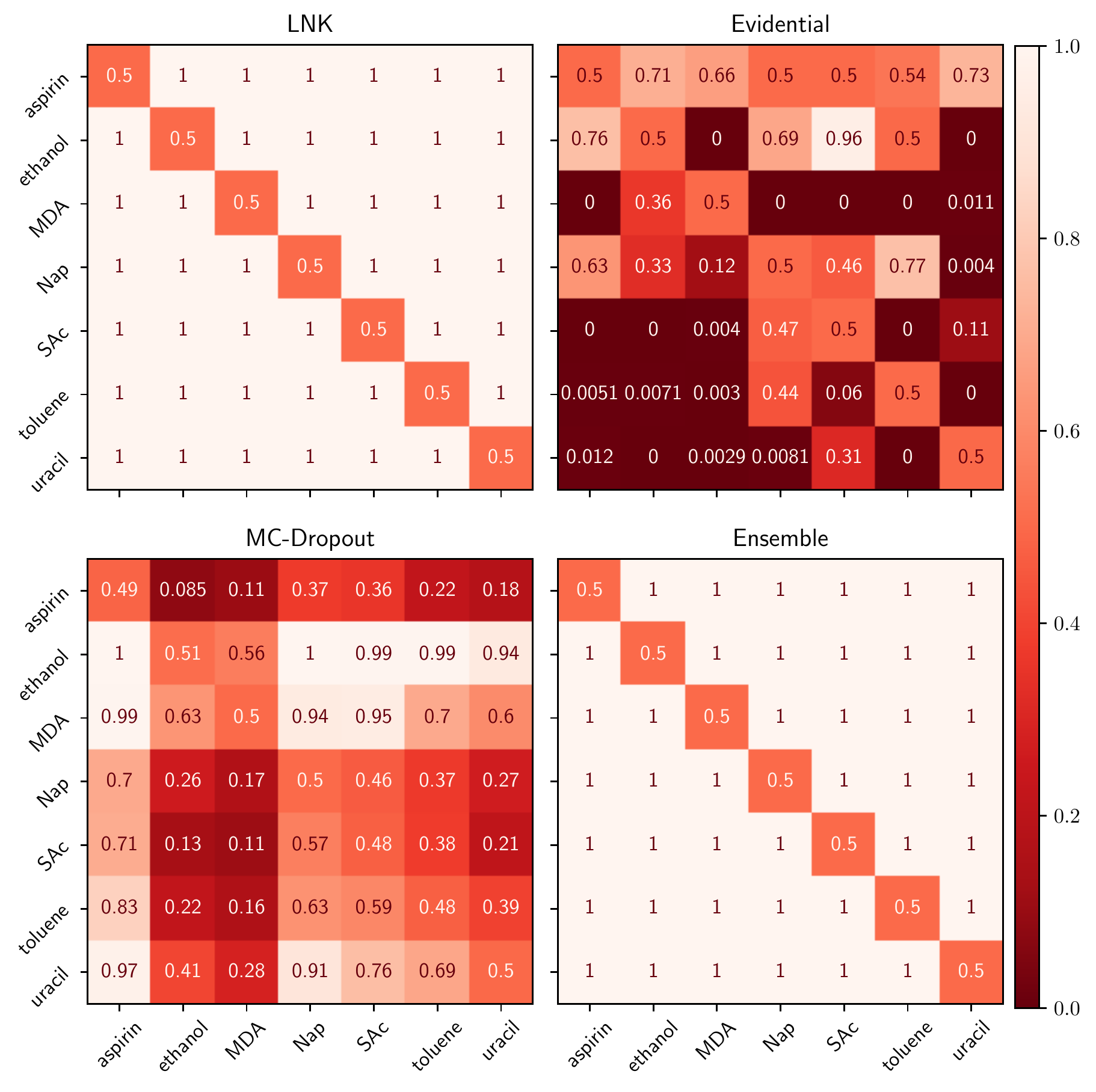}
\caption{Heatmap of AUC-ROC values from SchNet on MD17. Each row corresponds to a separate model trained on the molecule written on the left and tested on all other molecules.}\label{fig:md17_uq}
\end{figure}

In the following we test the performance of the methods on left-out class detection: we train a model on one of seven molecules and see whether it can distinguish the others based on its uncertainty estimate. An ideal estimator will have small values for the trained molecule and higher values for the other ones. This would enable a potential user to set a threshold to trust the model prediction whenever its uncertainty is lower than that threshold. To measure performance, we compute the area under the receiver operator curve of the uncertainty scores for ID and OOD data. A value closer to 1 indicates a better separation of the uncertainty scores, which means the model is better at detecting when it is operating on uncertain data.

For Evidential and \oursacro{} we use a single forward pass uncertainty estimator and for MCD and Ensembles we can use the variance of the energy prediction. In \autoref{fig:md17_uq}, we can see heat maps depicting the AUC-ROC score for each pairwise combination with SchNet as backbone. The rows show the molecule that the model is trained on and the off-diagonal columns are the respective OOD sample. On the diagonal, we expect a score of 0.5 while a score of 1 is optimal on the off-diagonals.

We observe that MCD's and Evidential's energy uncertainties are insufficient for such a task as both results exhibit no clear diagonal structure. Meanwhile, Ensembles exhibit perfect separation between in and out-of-distribution samples. With significantly fewer computational resources, \oursacro{} achieves identical results to Ensembles in a single forward pass, see \autoref{app:additional_experiments} for further results. This shows that both approaches can distinguish the molecule distribtions perfectly. 

We further evaluate the uncertainty estimation on QM7X, where we use the equilibrium structures to train our model and try to distinguish the out-of-equilibrium (OOE) ones. This task is more challenging as the distribution of the OOD samples is closer to the ID ones as OOE structures are generated by perturbing ID molecules. 

In \autoref{tab:qm7x_uq}, we can see the AUC-ROC scores with DimeNet++ as backbone. For MCD, we tested two dropout rates and neither could distinguish OOD samples. We observe similar behavior for the Evidential model.
In contrast, Ensembles again achieve close to perfect separation.
\oursacro{} is located in between these two extremes with an AUC-ROC score of 0.803, i.e., 2.1 times lower error than Evidential. Note, that this is obtained with a forward pass at inference, compared to five from the Ensemble. 
\begin{table}
    \centering
    \caption{OOD detection on QM7X based on the energy prediction with DimeNet$++$ backbone.}
    \vspace{0.05in}
    \resizebox{\columnwidth}{!}{
    \input{tables/qm7x/qm7x_dimenet_uq.tex}

    }
    \label{tab:qm7x_uq}
\end{table}
\subsection{\desseven{}}
Evidential and \oursacro{} both only require one single forward pass\footnote{With the backpropagation to calculate the forces} while MCD and Ensembles use 5 in our experiments. Thus it is natural that these techniques offer a higher \desseven{}. We evaluate the average time per sample for DimeNet++ on QM7X in \autoref{tab:speed}. We can see that both Evidential and \oursacro{} are indeed  more than $4.6\times$ faster than the variance based approaches. We further observe that \oursacro{} and Evidential are having very similar runtime which is due to similar number of parameters, as we in \cref{app:complexity}. Lastly, Ensembles are not only the having the computational demand at inference, but we also have to train multiple models which is additional overhead, especially in an active learning setting where one has to retrain the model many times.
\begin{table}
    \centering
    \caption{Inference runtime in $\si\ms$ on QM7X data for each UE method.}
    \vspace{0.05in}
    \resizebox{\columnwidth}{!}{
    \input{tables/speed.tex}
    }
    \label{tab:speed}
\end{table}
\subsection{Force Uncertainty}
As one cannot obtain force uncertainties with a single forward pass~\citep{gasteigerFastUncertaintyAwareDirectional2020}, force uncertainties directly violate our \desseven{} desiderata.
Nontheless, to provide a broad overview we here look at force uncertainties for OOE detection.
As single-forward pass methods cannot model $u_{\forces}$, we use the MCD approach on LNK by leveraging the dropout at inference time, denoted as MCD-\oursacro{}. We compare against Ensembles and MCD. For all methods we obtain the force uncertainties by computing the trace of the empirical covariance matrix.

\autoref{tab:force} lists the predictive performances and AUC-ROC scores for different dropout rates.
We observe that MCD-\oursacro{} consistently outperforms MCD both in predictive performance as well as in OOE detection.
Further, as the error increases significantly for MCD with increasing dropout rates, we find MCD-\oursacro{} to scale stably with clear improvements in OOE detection.
While Ensembles report close to perfect separation, it is also the computationally most intensive method due to its multiple trainings.

\begin{table}[htb]
    \centering
    \caption{Comparison of Ensembles, MCD, and \oursacro{} on QM7X with DimeNet++ as backbone for different dropout rates.}
    \vspace{0.05in}
    \resizebox{\columnwidth}{!}{
    \input{tables/qm7x/dropout_exp_lnk_dimenet.tex}
    }
    \label{tab:force}
\end{table}

%% file: tables/md17/schnet.tex
\begin{tabular}{llccc||c|c}
\toprule
{} & & MCD & Evidential & fixed-\oursacro{} & Backbone & Ensemble \\
\midrule
\multirow{3}{*}{aspirin} & Energy & 2.336 & 1.007 & \textbf{0.325} & 0.321 & 0.272\\ & Forces & 2.055 & 1.088 & \textbf{0.865} & 0.852 & 0.701\\ & Calibration & 1.256 & \textbf{1.077} & 1.370 & - & 1.354\\
\hline
\multirow{3}{*}{ethanol} & Energy & 0.646 & 0.081 & \textbf{0.065} & 0.064 & 0.06\\ & Forces & 0.925 & 0.328 & \textbf{0.271} & 0.112 & 0.191\\ & Calibration & \textbf{1.288} & 1.643 & 1.566 & - & 1.469\\
\hline
\multirow{3}{*}{MDA} & Energy & 1.493 & 0.130 & \textbf{0.111} & 0.111 & 0.098\\ & Forces & 1.467 & 0.535 & \textbf{0.481} & 0.461 & 0.368\\ & Calibration & \textbf{1.257} & 1.580 & 1.475 & - & 1.442\\
\hline
\multirow{3}{*}{Nap} & Energy & 2.739 & 0.346 & \textbf{0.129} & 0.131 & 0.124\\ & Forces & 1.385 & 0.357 & \textbf{0.266} & 0.255 & 0.211\\ & Calibration & 1.280 & \textbf{1.258} & 1.427 & - & 1.510\\
\hline
\multirow{3}{*}{SAC} & Energy & 2.301 & 0.165 & \textbf{0.142} & 0.141 & 0.132\\ & Forces & 1.853 & 0.607  & \textbf{0.481} & 0.465 & 0.378\\ & Calibration & \textbf{1.266} & 1.527 & 1.504 & - & 1.483\\
\hline
\multirow{3}{*}{toluene} & Energy & 1.662 & 0.202 & \textbf{0.106} & 0.108 & 0.099\\ & Forces & 1.404 & 0.402 & \textbf{0.310} & 0.296 & 0.230\\ & Calibration & \textbf{1.269} & 1.472 & 1.501 & - & 1.526\\
\hline
\multirow{3}{*}{uracil} & Energy & 1.759 & 0.168 & \textbf{0.121} & 0.119 & 0.115\\ & Forces & 1.945 & 0.480 & \textbf{0.338} & 0.326 & 0.265\\ & Calibration & \textbf{1.262} & 1.524 & 1.440 & - & 1.543\\
\bottomrule
\end{tabular}

%% file: tables/md17/average_comparison.tex
\begin{tabular}{llccc|c|c}
\toprule
   {} & {} & Dropout & Evidential & fixed-\oursacro{} & Backbone & Ensenmble \\
   \midrule
\multirow{2}{*}{DimeNet++} & Energy & 1.126 & 0.104 & \textbf{0.106} & 0.109 & 0.144\\ & Forces & 0.624 & 0.2 & \textbf{0.196} & 0.217 & 0.219\\ %
\hline
\multirow{2}{*}{SchNet} & Energy & 1.848 & 0.3 & \textbf{0.13} & 0.142 & 0.129\\ & Forces & 1.576 & 0.542 & 0.423 & 0.395 & 0.335\\
\hline
\multirow{2}{*}{NequIP} & Energy & 1.557 & \textbf{0.098} & 0.102 & 0.101 & 0.098\\ & Forces & 0.655 & 0.405 & \textbf{0.342} & 0.147 & 0.106\\

\bottomrule
\end{tabular}

%% file: tables/qm7x/qm7x_dimenet_predictive.tex
\begin{tabular}{lccccc|c}
\toprule
{} & MCD 20\% & MCD 1\% & Evidential & fixed-\oursacro{} & \oursacro{} & Ensemble\\
\midrule
Energy  &   0.24 & 0.09 & \textbf{0.02} & 0.046 & \textbf{0.0208} & 0.0523\\
Forces  &   0.0087 & \textbf{0.0043} & 0.0225 & 0.047 & 0.0046 & 0.0034\\
\midrule
Calibration    &  \textbf{1.279} & 1.285 & 1.629 & 1.293 & 1.339 & 2.597 \\
\bottomrule
\end{tabular}

%% file: tables/qm7x/qm7x_dimenet_uq.tex
\begin{tabular}{lccccc|c}
\toprule
{} & MCD 20\% & MCD 1\% & Evidential & fixed-\oursacro{} & \oursacro{} & Ensemble\\
\midrule
AUC-ROC & 0.511 & 0.504 & 0.587 & 0.742 & \textbf{0.803} & 0.947\\
\bottomrule
\end{tabular}

%% file: tables/speed.tex
\begin{tabular}{lcccc}
\toprule
{} & \oursacro{} & Evidential & MCD & Ensemble\\
\midrule
Runtime [ms]  & $59.3 \pm 2.68$ & $58.8 \pm 5.35$ & $229 \pm 0.25$ & $274 \pm 8.25$\\
\#Trainings         & 1 & 1 & 1 & 5\\
\bottomrule
\end{tabular}

%% file: tables/qm7x/dropout_exp_lnk_dimenet.tex
\begin{tabular}{llccccc}
\toprule
{Method} &  {Metric} & 0\% & 1\% & 5\% & 10\% & 20\%  \\
\midrule
\multirow{4}{*}{MCD} & Energy MAE & -- & 0.09 & 0.134 & 0.259 & 0.24  \\
& Forces MAE & -- & 0.004 & 0.005 & 0.012 & 0.009  \\
& AUC-ROC energy & -- &  0.504 & 0.511 & 0.55 & 0.511 \\
& AUC-ROC forces & -- &  0.86 & 0.87 & 0.89 & 0.91  \\
\midrule
\multirow{4}{*}{MCD-\oursacro{}} & Energy MAE & -- & 0.059 & 0.024 & 0.039 & 0.036\\
& Forces MAE & -- & 0.007 & 0.004 & 0.006 & 0.007\\
& AUC-ROC energy & -- & 0.6 & 0.69 & 0.72 & 0.72\\
& AUC-ROC forces & -- & 0.79 & 0.82 & 0.84 & 0.9\\
\midrule
\multirow{4}{*}{Ensemble} & Energy MAE & 0.052 & -- & -- & -- & -- \\
& Forces MAE & 0.003  & -- & -- & -- & --\\
& AUC-ROC energy & 0.947 & -- & -- & -- & --\\
& AUC-ROC forces & 0.996 & -- & -- & -- & -- \\
\bottomrule
\end{tabular}

%% file: include/conclusion.tex
\subsection*{Limitations}\label{sec:limitations}
Despite \oursacro{} meeting our desiderata, some aspects remain unexplored.
Firstly, while we cover several common techniques for UE for molecules this is not an exhaustive list, e.g., Bayesian neural networks~\citep{lambBayesianGraphNeural2020} or normalizing flows~\citep{kohlerSmoothNormalizingFlows2021}.
Secondly, recent research \cite{fu2022forces} has shown a gap between force prediction and simulation performance, emphasizing the importance of stability for backbones. However, due to noise in active learning \cite{mittal2019parting, ren2021survey} and challenges in bridging this gap for many core architectures \cite{fu2022forces}, we leave a simulation benchmark for UE as future work.

\section{Conclusion}\label{sec:conclusion}
In this work, we approached the field of uncertainty estimation (UE) for molecular force fields by defining six key desiderata, three `physics-informed' and three `application-focused', that UE for force fields should satisfy.
In a survey, we analyzed previous works based on these desiderata and found that none of them satisfies all of them.
To fill the gap, we proposed \oursacro{}, a localized GP-based extension to existing GNNs.
In our experimental evaluation, we found \oursacro{} to fulfill all desiderata and provide significant empirical improvements over previous single-forward methods for UE such as Monte Carlo dropout or evidential regression.

\section*{Acknowledgements}
This project is supported by the German Federal Ministry of Education and Research (BMBF), grant no. 01IS18036B, and the Free State of Bavaria under the Excellence Strategy of the Federal Government and the Länder. This project is further supported by the DAAD programme Konrad Zuse Schools of Excellence in Artificial Intelligence, sponsored by the Federal Ministry of Education and Research.

%% file: include/appendix.tex
\section{Implementation details}\label{app:implementation_details}
\subsection{\ours{}}
To avoid constant cross-referencing, we recap essential parts of Section \ref{sec:method} here. Our model predicts a distribution for the energy value $\energy_*$ as: 
\begin{align}
    p(E_\star \vert \pos, \inputfeatures) &= \sum_i^\natoms \gp_\phi \circ h_{\text{rep}}(\pos, \inputfeatures)_{i}\\ 
    &\sim \normal\left(E_\star \,\vert\,\sum_i^\natoms \EX [\energy_\star]_i, \sum_{ij}^\natoms \Cov(\energy_\star)_{ij}\right).\nonumber
\end{align}
Here $h_{\text{rep}}$ is a GNN backbone while $\gp_\phi$ is the gaussian process with parameters $\phi$ and $n$ is the number of atoms of the molecule. Any GNN that implements atom-wise embeddings qualifies as backbone. However, to fulfill the desiderata we use recent approaches for ML force fields that implement \desone{} and \destwo{}. We show the schematic depiction of our model in Figure \ref{fig:model_figure}. For one molecule one obtains $n$ different embeddings. For each we use the same GP to predict an energy distribution. The final predictive distribution is the sum of the individual distributions. 

As we use an approximate GP with inducing points we have the following additional hyperparameters we consider: number of inducing points, inducing point initialization and kernel lengthscale initialization. For the mean value of the GP prior we used a standard learn-able constant. For the number of inducing points we tested multiple different values in $\{10, ,25, 50, 100, 256, 512\}$. We choose 100 for MD17 as the training set only consists of 1000 data points and more inducing points lead to overfitting. For QM7X we used 256 as we have more data points. 

For the inducing point initialization we tested three different options: first, we use randomly generated points. Secondly, we use the k-means algorithm to find multiple cluster centers within the training embeddings and choose these. Lastly, we take the first $\lfloor N_{\text{ind pnts}} / N_{\text{atoms}}\rfloor$ training embeddings for the atom embeddings. As we noticed no significant difference between the second and the third we use the last one as it is faster to compute. 

For the fixed-LNK a crucial hyperparameter is the lengthscale initialization. As the inducing points might be far apart in a high-dimensional embedding space, such as for DimeNet++ with 256 dimensions, a too small initial lengthscale leads to sparse gradients and poor to no optimization of the GP posterior. Thus, we use the mean distance within all clusters of inducing points which are $N_{\text{atoms}}$ many. If we have varying number of atoms it will $N_{\text{atoms}}$ refers to the maximum number of atoms of a molecule in the training dataset. 

\begin{table}[!h]
    \centering
    \input{tables/lnk_details}
    \caption{Caption}
    \label{tab:lnk_hyperparameters}
\end{table}

\subsection{Baseline details}\label{app:baselinedetails}
\textbf{Backbones.} We use three different model architectures: DimeNet++ \citep{klicpera2020Fast}, SchNet \citep{schuttSchNetDeepLearning2018}, and NequIP \cite{batznerEquivariantGraphNeural2022}. Unless otherwise stated, we use the default hyperparameters of each model architecture and training as stated in the original works. For DimeNet++ and SchNet, specific hyperparameters used for training are listed in Table \ref{tab:training_details}. For NequIP we used exactly the same hyperparameters as the original implementation and we omit the details here. The hyperparameters specific to the datasets and that are used with all models are listed in Table \ref{tab:data_hyperparameters}. Unless otherwise specified, we train the backbones the standard combined energy and force loss~\citep{gasteigerDirectionalMessagePassing2019}:

\begin{align}
    \mathcal{L}_\text{total} =& (1-\rho)\mathcal{L}_\text{energy} + \rho\mathcal{L}_\text{force},\\
    \mathcal{L}_\text{energy} =& \frac{1}{\ndatapoints}\sum_{i=1}^\ndatapoints \vert f_\theta(\pos, \inputfeatures) - \energy^*\vert, \\
    \mathcal{L}_\text{force} =& \frac{1}{\ndatapoints}\sum_{i=1}^\ndatapoints \sum_{j=1}^{\natoms} \sqrt{\sum_{k=1}^3 \left( \frac{\partial}{\partial \pos_i}f_\theta(\pos_i, \inputfeatures_i) - \forces_i^*\right)_{jk}^2}.
\end{align}

\begin{table}[h!]
\centering
\caption{Hyperparameters of DimNet++ and SchNet}\label{tab:training_details}
\input{tables/training_details.tex}
\end{table}

\begin{table}[h!]
\centering
\caption{Hyperparameters of the datasets used with all models}\label{tab:data_hyperparameters}
\input{tables/data_hyperparams.tex}
\end{table}

\subsection{Monte Carlo Dropout: implementation details}
\textbf{DimeNet++.} This architecture only uses representations of dimension 0, so we use standard dropout. We apply it across different blocks after every fully-connected layer that uses an activation function. One exception is the directional message passing block, where we apply dropout after the two Hadamard product operations and once after the final fully-connected layer. Note that applying dropout separately on the two operands of the Hadamard product would result in effectively more dropped neurons than applying it after.

\textbf{SchNet.} Similar to DimeNet++,  we here also apply dropout after each linear layer that is followed by an activation function. Specifically, we use dropout inside the interaction, convolution, and output layers.

\textbf{NequIP.} This architecture uses higher representations of dimension 1. Therefore, in order to preserve the equivariance property, we apply dropout at every layer of the radial network inside the interaction block which uses only representations of dimension 0. We also drop representations of dimensions 1 by applying the MCD layer described above after each convolution layer.

\clearpage
\section{Complexity details}\label{app:complexity}
In \cref{tab:num_parameters} we show the number of parameters for each model. We used the default parameter settings of the respective paper. Notably, our model \oursacro{}, introduces only a marginal increase in additional parameters compared to existing models. As seen in \cref{tab:speed}, the slight increase in parameters translates to a minimal decrease in computational speed compared to the Evidential model.

We observe that Ensembles have five times more parameters than the other models. An interesting future work may be investigating whether the Ensembles can be reduced in their number of parameters to have the same computational demand as the single model. However, according to research conducted by \citet{batatia2022design}, a reduction in the number of channels on NequIP is reduced from 32 to 16 (which is only a reduction of about 2.6), resulted in $8.4\%$ decrease in force prediction performance and a $19.35\%$ decline in energy prediction performance. 

\begin{table}[h!]
    \centering
    \begin{tabular}{l|rrr}
\toprule
Approach & DimeNet++ & SchNet & NequIP\\
\midrule
MCD & 1,885,830 & 455,809 & 189,080\\
Evidential & 1,889,670 & 456,004 & 189,128\\
\oursacro{} & 1,920,150 & 468,263 & 197,567\\
Ensemble & 9,429,150 & 2,279,045 & 1,119,800\\
\bottomrule
    \end{tabular}
    \caption{Number of parameters for each UE approach and backbone. SchNet is the version of Pytorch-Geometric with 6 interaction blocks.}
    \label{tab:num_parameters}
\end{table}

\section{Additional Experiments} \label{app:additional_experiments}
In the following we list additional experiments, i.e., evaluating different backbones or approaches.

\subsection{\oursacro{} vs. SVGP-DKL}\label{app:svgp_vs_lnk}
To analyze the improvement from the localization of \oursacro{}, we compare the performance of \oursacro{} against SVGP-DKL on MD17 and QM7X. The results are listed in \cref{tab:lnk_vs_svgpdkl}. For MD17 We observe that \oursacro{} decreases the error by $5\%$ on average for the energy prediction and by $28\%$ for forces. For QM7X we decrease the error by $99.3\%$ for energy but increase the forces slightly by $6.5\%$.
\begin{table}[h!]
    \centering
    \input{tables/svgpdkl_vs_lnk}
\caption{Comparison of LNK vs SVGP-DKL with DimeNet++ as backbone.}
\label{tab:lnk_vs_svgpdkl}
\end{table}

\subsection{Graph Mixture Density Network}\label{app:gmdn}
In the following we investigate the performance of the Graph Mixture Density Network (GMDN) \cite{errica2021graph}. This network outputs a mixture of distributions as resulting predictive distribution. Each individual distribution is learned by a neural network head. Here we investigate \desfive{} and \dessix{}. In \cref{app:locality_consideratins}, we show that the network is not necessarily fulfilling \desfour{}.

In \cref{tab:gmdn_vs_lnk}, we observe that \oursacro{} is outperforming GMDN on every single target. For aspirin and napthalene the model did not learn to predict the energy properly. As GMDN was not tested for molecular property prediction by the authors, trained for different hyperparamters for the number of hidden units, the number of mixture components and the number of output layers, see \cref{app:locality_consideratins} for a detailed model description. For each target we used the best performing setting, i.e., not a fix hyperparameter setting for all molecules on MD17 as done for all other baselines but the best set for each target. However, there might be hyperparameter sets where this network yields better results and we leave it for future work to adapt GMDN to this setting.

\begin{figure}[h!]
    \centering
    \includegraphics[width=.3\textwidth]{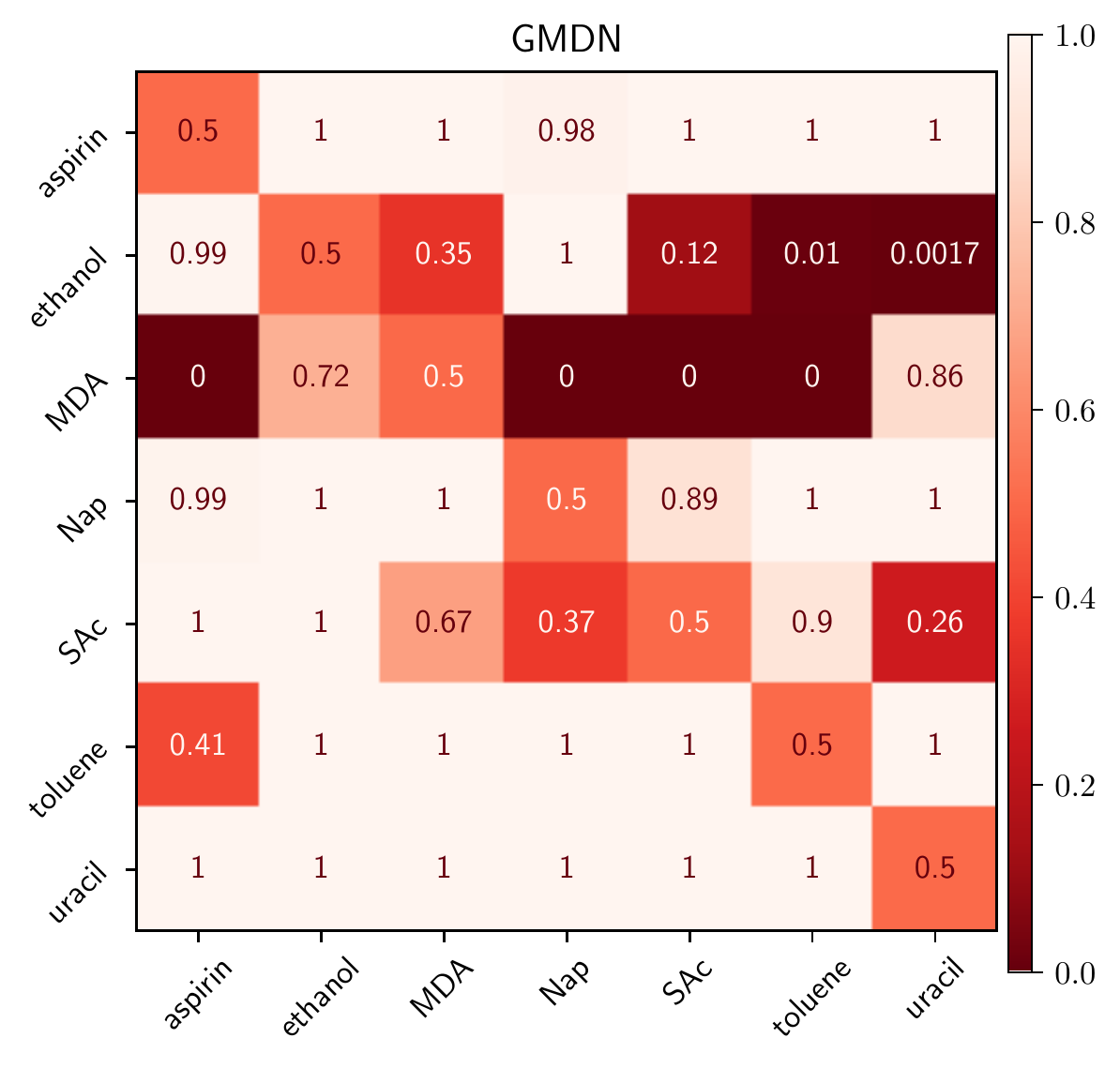}
    \caption{OOD detection with GMDN and SchNet as backbone model.}
    \label{fig:gmdn_md17_ood}
\end{figure}

In \cref{fig:gmdn_md17_ood}, we see the OOD detection performance we use to measure \dessix{}. Compared to MCD or Evidential, the method has higher separation for most of the OOD molecules. However, it is not close to the results of Ensemble or \oursacro{}.

\begin{table}[h!]
    \centering
    \input{tables/gmdn}
\caption{Comparison of LNK vs GMDN with SchNet as backbone. For MD17 We observe that \oursacro{} outperforms GMDN on every target.}
\label{tab:gmdn_vs_lnk}
\end{table}

\subsection{Additional experiments on MD17}
In the following we list the experimental results for the other backbones. We use either 69 samples or 5 samples for MCD evaluation. 

\begin{table}[h!]
\centering
\caption{MAE of Dimenet++ on MD17 with different uncertainty estimation methods. MCD stands for MC-Dropout (energies in $\si{\kcal\per\mole}$, forces in $\si{\kcal\per\mole\per\angstrom}$)}
\include{tables/md17/dimenet_combined.tex}
\end{table}

\begin{table}[h!]
\centering
\caption{MAE of NequIP on MD17 with different uncertainty estimation methods (energies in $\si{\kcal\per\mole}$, forces in $\si{\kcal\per\mole\per\angstrom}$)}
\include{tables/md17/nequip_combined.tex}
\end{table}

\begin{table}[h!]
\centering
\caption{MAE of SchNet on MD17 with different uncertainty estimation methods (energies in $\si{\kcal\per\mole}$, forces in $\si{\kcal\per\mole\per\angstrom}$)}\label{tab:schnet_more_samples}
\include{tables/md17/schnet_combined.tex}
\end{table}

\begin{figure}[h!]
\centering
\includegraphics[width=1.0\textwidth]{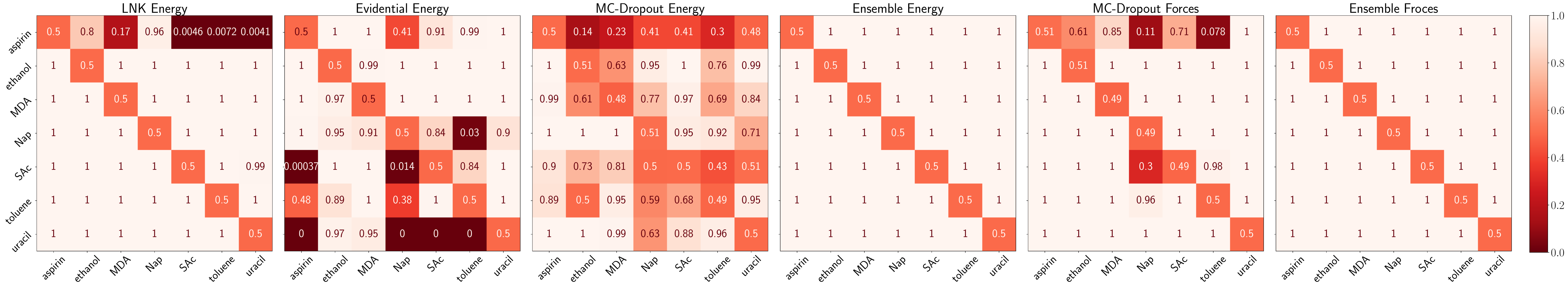}
\caption{Heatmap of AUC-ROC on MD17 values based on different UE methods with DimeNet++ as backbone. Each row corresponds to a separate model trained on the molecule written on the left and tested on all other molecules.}
\end{figure}

\begin{figure}[h!]
\centering
\includegraphics[width=1.0\textwidth]{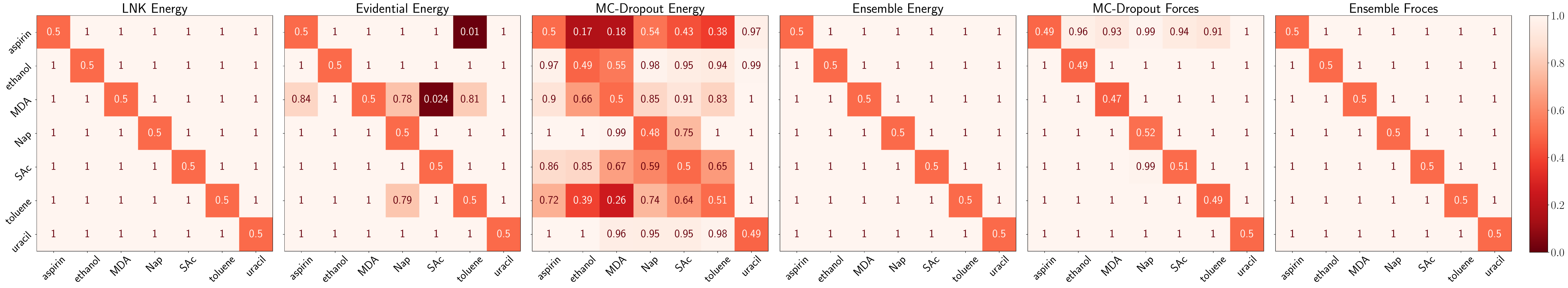}
\caption{Heatmap of AUC-ROC on MD17 values based on different UE methods with NequIP as backbone. Each row corresponds to a separate model trained on the molecule written on the left and tested on all other molecules.}
\end{figure}

\subsection{Additional experiments on QM7X}
In the following we show the experimental results for QM7X for other backbones and for dropout with 5 and 69 samples respectively.

\begin{table}[h!]%
    \caption{MAE of Dimenet++ trained on QM7X-Equilibrium with different uncertainty estimation methods (energies in $\si{\eV}$, forces in $\si{\eV\per\angstrom}$)}
    \include{tables/qm7x/qm7x_dimenet.tex}
\end{table}
\begin{table}[h!]%
\caption{MAE of Dimenet++ ($N=69$ samples) trained on QM7X-Equilibrium with different uncertainty estimation methods (energies in $\si{\eV}$, forces in $\si{\eV\per\angstrom}$)}
    \include{tables/qm7x/qm7x_dimenet_69.tex}
\end{table}

\begin{table}[h]
\centering
\caption{MAE of SchNet, and NequIP trained on QM7X-Equilibrium with different uncertainty estimation methods (energies in $\si{\eV}$, forces in $\si{\eV\per\angstrom}$). The numbers marked with a star denote that they were  obtained after removing 7 out of the 1537 test molecules that were causing numerical instability of the NequIP model.}
\include{tables/qm7x/qm7x_other_models.tex}
\end{table}
\begin{table}[h]
\centering
\caption{MAE of SchNet, and NequIP ($N=69$ samples) trained on QM7X-Equilibrium with different uncertainty estimation methods (energies in $\si{\eV}$, forces in $\si{\eV\per\angstrom}$). The numbers marked with a star denote that they were  obtained after removing 7 out of the 1537 test molecules that were causing numerical instability of the NequIP model.}
\include{tables/qm7x/qm7x_other_models_69samples.tex}
\end{table}

\subsection{Dropout Rate Experiment}\label{app:dropout_exp}
We compare different dropout probabilities for MCD on QM7X for DimeNet++ and SchNet. 
\begin{table}[h]
\centering
\caption{MAE and uncertainty metrics of MC-Dropout Dimenet++ ($N=5$ samples) trained on QM7X-Equilibrium with different dropout rates (energies in $\si{\eV}$, forces in $\si{\eV\per\angstrom}$)}
\include{tables/qm7x/dropout_exp_dimenet.tex}
\end{table}
\begin{table}[h]
\centering
\caption{MAE and uncertainty metrics of MC-Dropout SchNet ($N=5$ samples) trained on QM7X-Equilibrium with different dropout rates (energies in $\si{\eV}$, forces in $\si{\eV\per\angstrom}$)}
\include{tables/qm7x/dropout_exp_schnet.tex}
\end{table}

\clearpage
\section{\desfour{} consideration of baseline methods} \label{app:locality_consideratins}
\textbf{Ensemble}
As written above, we estimate the uncertainty of a molecule for an ensemble ${f_1, \dots, f_5}$ of five models $f_i$ as:
\begin{equation}
    \mathbf{u}_E(\molecule) = \sqrt{\text{Var}\left(f_1(\molecule), \dots, f_N(\molecule)\right)}.
\end{equation}
If we duplicate the molecule and have it sufficiently far apart, then we get: 
\begin{equation}
    f_i(\{\molecule, \molecule\}) = 2 * f_i(\molecule),
\end{equation}
as the embeddings of the respective molecules do not influence each other and the final result is the sum over the atom predictions. Hence, we can derive that the uncertainty estimate of the ensemble is satisfies \desfour{}:
\begin{equation}\begin{aligned}
    \uncertainty_{\energy}(\{\molecule, \molecule\}) &= \sqrt{\variance\left(f_1(\{\molecule, \molecule\}), \dots, f_N(\{\molecule, \molecule\})\right)}\\ &= \sqrt{\variance\left(2f_1(\molecule), \dots, 2f_N(\molecule)\right)}\\ &= \sqrt{4 \cdot \text{Var}\left(f_1(\molecule), \dots, f_N(\molecule)\right)}\\ &= 2\uncertainty_{\energy}(\molecule).
\end{aligned}\end{equation}

\textbf{Graph Mixture Density Networks} \cite{errica2021graph} can be formulated as: 
\begin{equation}
    \sum_{i=1}^CP(y_g| Q_g^i, g) P(Q_g^i|g)
\end{equation}

where C is the number of mixing components (experts), $g$ is the graph, $Q_g^i$ is the mixing weight $i$ for graph $g$. It is modelled by a network $\Phi_i$. $y_g$ is the target we want to model, i.e. $P(y_g | g)$.  Given the node embeddings

\begin{equation}
    \mathbf h_v^{l+1} = \phi^{l+1}\left(\mathbf h_v^l, \Psi(\{\psi^{l+1}(\mathbf h_u^l) | u \in \mathcal N_v\})\right), \forall l = 1 \dots L
\end{equation}

The final node representation is given by $\mathbf h_v :=  \mathbf h_v^L$. $P(Q_g^i|g)$ models a mixing weight by a neural network. The interesting part is the emission $P(y_g| Q_g^i, g)$, i.e. the distribution parameters. They are compute as: 

\begin{equation}
\mathbf \mu_i, \mathbf \Sigma_i = \Phi_i(\mathbf h_g) = f_i(r_g^i(\mathbf h_{\mathcal V_g})).
\end{equation}

Here, $\mathbf h_g$ is the aggregated embedding for the full graph, $\Phi_i$ is a sub-network to compute the individual distributional parameter, $f_i$ is a neural network, $r_g^i$ is the readout network and $\mathcal V_g$ is the set of nodes of graph g.

If we now construct a graph $\hat g := \{g, g\}$ whithout edges between the individual graphs $g$, then, due to the non-linearity of $f_i$ (or $\Phi_i$), we would have:

\begin{equation}
\Phi_i(\mathbf h_{\hat g}) = \hat \mu_i, \hat{\mathbf \Sigma}_i \neq 2\mu_i, 2\Sigma_i = \Phi_i(\mathbf h_g).
\end{equation}
Hence, GMDN is not necessarily fulfilling \desfour{}.

\clearpage
\section{Metrics}
\subsection{Calibration}
We use the calibration score introduced by \citet{accurate-uncertainties-deep-learning-regression}. It computes the difference between the percentage of targets lying in a certain confidence region and the corresponding percentile p. In mathematical notation, the calibration metric is: 

\begin{equation}
    p_{pred} = \frac{1}{N} \sum_i \mathbb{I}\left[\mathbb{P}\left(y \leq y^{*, (i)}| \theta^{(i)}\right) \in I_p\right],
\end{equation}

where $I_p = [0, \frac{p}{2}] \cup [1-\frac{p}{2}, 1]$. To obtain a single score we compute the norm:
\begin{equation}
    \text{s}_{\text{calib}} := \sqrt{\sum\nolimits_p (p - p_{\text{pred}})^2}\;\; , \quad p \in \{0.1, \dots, 0.9\}
\end{equation}

\subsection{OOD Detection}
The OOD detection task can be viewed as a binary classification task. We assign class 1 to ID data and class 0 to OOD data and use the predicted uncertainty values as scores, based on which we want to distinguish between ID and OOD data. This enables to compute UE metrics using, for instance, the area under the receiver operating characteristic curve (AUC-ROC). We obtain numbers in [0,1], where higher values indicate better performance. For the MC-Dropout and the ensemble methods, we use the variance of the energy samples as the uncertainty estimate corresponding to the energy prediction and the trace of the sample covariance matrix of the forces samples as the uncertainty estimate corresponding to the forces prediction. We found that the trace is more stable than taking the determinant and leads to good results. For the evidential models, we use the analytic variance of the energy prediction, which we can compute in closed form based on the parameters of the evidential distribution, as the uncertainty estimate corresponding to the energy prediction. We don't use any uncertainty estimates for the forces prediction, because there is no obvious way how to do that.

%% file: tables/lnk_details.tex
\begin{tabular}{llll}
\hline
Model & Hyperparameter & QM7-X & MD17 \\
\hline
\multirow{4}{*}{Fixed-LNK}          
& learning rate                     &  0.01         &  0.01 \\
& patience (epochs)                 & 50            & 50\\
& force weighting factor $\rho$     & 0.99          & 0.99\\
& \# inducing points                & 256           & 100\\
& inducing point initialization     & first k       & first k\\\hline
\multirow{4}{*}{LNK}        
& learning rate                     & 0.001         & -\\
& patience (epochs)                 & 50            & -\\
& force weighting factor $\rho$     & 0.99          & -\\
& \# inducing points                & 256           & -\\
& inducing point initialization     & first k       & -\\\hline
\end{tabular}

%% file: tables/training_details.tex
\begin{tabular}{llll}
\toprule
Model                                 & Hyperparameter                             & QM7-X                                                    & MD17                                                    \\ \midrule
\multirow{8}{*}{DimeNet++ Dropout}    & learning rate                              & 0.002                                                    & 0.001                                                   \\
                                      & warmup steps                               & 550                                                      & 10000                                                   \\
                                      & decay steps                                & 6600000                                                  & 1200000                                                 \\
                                      & decay rate                                 & 0.01                                                     & 0.01                                                    \\
                                      & EMA decay                                  & 0.999                                                    & 0.999                                                   \\
                                      & patience (epochs)                          & 50                                                       & 50                                                      \\
                                      & force weighting factor $\rho$              & 0.99                                                     & 0.99                                                    \\
                                      & dropout locations                          & \multicolumn{2}{c}{\begin{tabular}[c]{@{}c@{}}embedding block, \\ interaction block, \\ output block\end{tabular}} \\ \midrule
\multirow{8}{*}{DimeNet++ Evidential} & learning rate                              & 0.001                                                    & 0.001                                                   \\
                                      & warmup steps                               & 550                                                      & 10000                                                   \\
                                      & decay steps                                & 660000                                                   & 1200000                                                 \\
                                      & decay rate                                 & 0.01                                                     & 0.01                                                    \\
                                      & EMA decay                                  & 0.99                                                     & 0.999                                                   \\
                                      & patience (epochs)                          & 50                                                       & 50                                                      \\
                                      & force weighting factor $\rho$              & 0.9                                                      & 0.9                                                     \\
                                      & evidential loss weighting factor $\lambda$ & 0.2                                                      & 0.2                                                     \\ \midrule
\multirow{8}{*}{SchNet Dropout}    & learning rate                              & 0.001                                                     & 0.001                                                    \\
                                   & warmup steps                               & 1000                                                      & 10000                                                    \\
                                   & decay steps                                & 4000000                                                   & 4000000                                                  \\
                                   & decay rate                                 & 0.01                                                      & 0.01                                                     \\
                                   & EMA decay                                  & 0.999                                                     & 0.999                                                    \\
                                   & patience (epochs)                          & 50                                                        & 50                                                       \\
                                   & force weighting factor $\rho$              & 0.99                                                      & 0.99                                                     \\
                                   & dropout locations                          & \multicolumn{2}{c}{\begin{tabular}[c]{@{}c@{}}interaction block,\\ convolution layers, \\ output block\end{tabular}} \\ \midrule
\multirow{8}{*}{SchNet Evidential} & learning rate                              & 0.001                                                     & 0.001                                                    \\
                                   & warmup steps                               & 1000                                                      & 1000                                                     \\
                                   & decay steps                                & 4000000                                                   & 4000000                                                  \\
                                   & decay rate                                 & 0.01                                                      & 0.01                                                     \\
                                   & EMA decay                                  & 0.999                                                     & 0.999                                                    \\
                                   & patience (epochs)                          & 50                                                        & 50                                                       \\
                                   & force weighting factor $\rho$              & 0.9                                                       & 0.9                                                      \\
                                   & evidential loss weighting factor $\lambda$ & 0.2                                                       & 0.2                                                      \\ \bottomrule
\end{tabular}

%% file: tables/data_hyperparams.tex
\begin{tabular}{lll}
\toprule
                 & QM7-X & MD17 \\ \midrule
batch size       & 32    & 1    \\
train set size   & 33213 & 1000 \\
val set size     & 4151  & 1000 \\
standardize data & False & True \\ \bottomrule
\end{tabular}

%% file: tables/svgpdkl_vs_lnk.tex
\begin{tabular}{l|llcc}
\toprule
{Dataset} & Training Target & & \oursacro{} & SVGP-DKL\\
\midrule
\multirow{14}{*}{MD17} & \multirow{2}{*}{aspirin} & Energy & \textbf{0.145} & 0.208\\ & &Forces & \textbf{0.359} & 0.504\\
\cmidrule{2-5}
&\multirow{2}{*}{ethanol} & Energy & \textbf{0.054} & 0.065\\ && Forces & \textbf{0.201} & 0.296\\
\cmidrule{2-5}
&\multirow{2}{*}{malonaldehyde} & Energy & \textbf{0.096} & 0.107\\ && Forces & \textbf{0.299} & 0.478\\
\cmidrule{2-5}
&\multirow{2}{*}{napthalene} & Energy & 0.133 & \textbf{0.107}\\ && Forces & \textbf{0.121} & 0.166\\
\cmidrule{2-5}
&\multirow{2}{*}{salicylic acid} & Energy & \textbf{0.119} & \textbf{0.12}\\ && Forces & \textbf{0.257} & 0.324\\
\cmidrule{2-5}
&\multirow{2}{*}{toluene} & Energy & 0.094 & \textbf{0.091}\\ && Forces & \textbf{0.122} & 0.191\\
\cmidrule{2-5}
&\multirow{2}{*}{uracil} & Energy & \textbf{0.102} & 0.108\\ && Forces & \textbf{0.222} & 0.265\\
\hline
\multirow{2}{*}{QM7X} & \multirow{2}{*}{Equilibrium} & Energy & \textbf{0.05} & 7.047\\ && Forces & 0.049 & \textbf{0.046}\\
\bottomrule
\end{tabular}

%% file: tables/gmdn.tex
\begin{tabular}{l|llcc}
\toprule
{Dataset} & Training Target & & \oursacro{} & GMDN\\
\midrule
\multirow{14}{*}{MD17} & \multirow{2}{*}{aspirin} & Energy & \textbf{0.325} & 174.051\\ & &Forces & \textbf{0.865} & 0.907\\
\cmidrule{2-5}
&\multirow{2}{*}{ethanol} & Energy & \textbf{0.065} & 0.071\\ && Forces & \textbf{0.271} & 0.356\\
\cmidrule{2-5}
&\multirow{2}{*}{malonaldehyde} & Energy & \textbf{0.111} & 0.124\\ && Forces & \textbf{0.481} & 0.585\\
\cmidrule{2-5}
&\multirow{2}{*}{napthalene} & Energy & \textbf{0.129} & 85.265\\ && Forces & \textbf{0.266} & 0.316\\
\cmidrule{2-5}
&\multirow{2}{*}{salicylic acid} & Energy & \textbf{0.142} & 0.160\\ && Forces & \textbf{0.481} & 0.605\\
\cmidrule{2-5}
&\multirow{2}{*}{toluene} & Energy & \textbf{0.106} & 0.112\\ && Forces & \textbf{0.310} & 0.381\\
\cmidrule{2-5}
&\multirow{2}{*}{uracil} & Energy & \textbf{0.121} & 0.129\\ && Forces & \textbf{0.338} & 0.476\\
\bottomrule
\end{tabular}

%% file: tables/md17/dimenet_combined.tex
\begin{tabular}{llcccc|c|c}
\toprule
{} & & MCD ($N=5$) & MCD ($N=69$) & Evidential & \oursacro{} & Backbone & Ensemble\\
\midrule
\multirow{3}{*}{aspirin} & Energy & 1.243 & 0.445 & \textbf{0.165} & 0.175 & 0.174 & 0.155\\ & Forces & 0.793 & 0.633 & 0.388 & \textbf{0.376} & 0.356 &0.281\\ & Calibration & 1.263 & 1.517 & \textbf{0.480} & 1.686 & - & 1.454\\
\midrule
\multirow{3}{*}{ethanol} & Energy & 0.892 & 0.280 & \textbf{0.054} & \textbf{0.055} & 0.055 & 0.054\\ & Forces & 0.360 & 0.260 & \textbf{0.154} & 0.186 & 0.173 & 0.137\\ & Calibration & \textbf{1.282} & 1.572 & 1.487 & 1.688 & - & 1.528\\
\midrule
\multirow{3}{*}{malonaldehyde} & Energy & 1.147 & 0.373 & \textbf{0.080} & 0.093 & 0.095 & 0.09\\ & Forces & 0.632 & 0.463 & \textbf{0.254} & 0.336 & 0.326 & 0.269\\ & Calibration & \textbf{1.287} & 1.563 & 1.500 & 1.680 & - & 1.514\\
\midrule
\multirow{3}{*}{napthalene} & Energy  & 1.078 & 0.338 & \textbf{0.117} & \textbf{0.117} & 0.114 & 0.117\\ & Forces & 0.536 & 0.412 & \textbf{0.098} & 0.114 & 0.101 & 0.075\\ & Calibration & \textbf{1.283} & 1.555 & 1.512 & 1.664 & - & 1.613\\
\midrule
\multirow{3}{*}{salicylic acid} & Energy & 1.215 & 0.392 & \textbf{0.110} & 0.121 & 0.12 & 0.113\\ & Forces & 0.762 & 0.585 & \textbf{0.220} & 0.274 & 0.243 & 0.19\\ & Calibration & 1.299 & 1.561 & \textbf{0.699} & 1.687 & - & 1.562\\
\midrule
\multirow{3}{*}{toluene} & Energy & 1.09 & 0.331 & 0.092 & \textbf{0.091} & 0.09 & 0.09\\ & Forces & 0.52 & 0.378 & \textbf{0.104} & 0.136 & 0.121 & 0.089\\ & Calibration & \textbf{1.287} & 1.571 & 1.582 & 1.683 & - & 1.582\\
\midrule
\multirow{3}{*}{uracil} & Energy & 1.219 & 0.399 & \textbf{0.110} & 0.112 & 0.116 & 0.387\\ & Forces & 0.762 & 0.577 & \textbf{0.158} & 0.223 & 0.196 & 0.49\\ & Calibration & 1.273 & 1.556 & \textbf{0.816} & 1.688 & - & 1.11\\
\bottomrule
\end{tabular}

%% file: tables/md17/nequip_combined.tex
\begin{tabular}{llcccc|c|c}
\toprule
{} & & MCD ($N=5$) & MCD ($N=69$) & Evidential & \oursacro{} & Backbone & Ensemble\\
\midrule
\multirow{3}{*}{aspirin} & Energy & 2.466 & 0.744 & 0.194 & \textbf{0.142} & 0.153 & 0.144\\ & Forces & 0.943 & 0.754 & 0.54 & \textbf{0.304} & 0.287 & 0.210\\ & Calibration & \textbf{1.278} & 1.553 & 1.688 & - & - & 1.472\\
\midrule
\multirow{3}{*}{ethanol} & Energy & 0.932 & 0.272 & 0.073 & \textbf{0.056} & 0.053 & 0.050\\ & Forces & 0.588 & 0.410 & 0.401 & \textbf{0.15} & 0.137 & 0.096\\ & Calibration & \textbf{1.283} & 1.576 & 1.692 & - & - & 1.518\\
\midrule
\multirow{3}{*}{malonaldehyde} & Energy & 1.131 & 0.432 & 0.103 & \textbf{0.086} & 0.079 & 0.075\\ & Forces & 0.733 & 0.535 & 0.496 & \textbf{0.307} & 0.234 & 0.166\\ & Calibration & \textbf{1.271} & 1.538 & 1.675 & - & - & 1.494\\
\midrule
\multirow{3}{*}{napthalene} & Energy & 1.615 & 0.456 & 0.066 & - & 0.114 & 0.115\\ & Forces & 0.445 & 0.296 & 0.3 & - & 0.054 & 0.042\\ & Calibration & \textbf{1.273} & 1.584 & 1.683 & - & - & 1.619\\
\midrule
\multirow{3}{*}{salicylic acid} & Energy & 1.843 & 0.561 & \textbf{0.11} & 0.127 & 0.109 & 0.107\\ & Forces & 0.750 & 0.560 & 0.433 & \textbf{0.27} & 0.146 & 0.103\\ & Calibration & \textbf{1.275} & 1.580 & 1.667 & - & - & 1.558\\
\midrule
\multirow{3}{*}{toluene} & Energy & 1.428 & 0.437 & \textbf{0.071} & 0.093 & 0.089 & 0.089\\ & Forces & 0.513 & 0.348 & 0.299 & \textbf{0.068} & 0.069 & 0.051\\ & Calibration & \textbf{1.272} & 1.594 & 1.689 & - & - & 1.611\\
\midrule
\multirow{3}{*}{uracil} & Energy & 1.485 & 0.449 & \textbf{0.069} & 0.109 & 0.107 & 0.104\\ & Forces & 0.611 & 0.407 & \textbf{0.363} & 0.955 & 0.101 & 0.073\\ & Calibration & \textbf{1.293} & 1.583 & 1.701 & - & - & 1.597\\
\bottomrule
\end{tabular}

%% file: tables/md17/schnet_combined.tex
\begin{tabular}{llcccc|c|c}
\toprule
{} & & MCD ($N=5$) &MCD ($N=69$) & Evidential & \oursacro{} & Backbone & Ensemble \\
\midrule
\multirow{3}{*}{aspirin} & Energy & 2.336 & 1.302 & 1.007 & \textbf{0.271} & 0.321 & 0.272\\ & Forces & 2.055 & 1.773 & 1.088 & \textbf{0.857} & 0.852 & 0.701\\ & Calibration & 1.256 & 1.419 & 1.077 & \textbf{0.78} & - & 2.275\\
\midrule
\multirow{3}{*}{ethanol} & Energy & 0.646 & 0.284 & 0.081 & \textbf{0.066} & 0.064 & 0.06\\ & Forces & 0.925 & 0.697 & 0.328 & \textbf{0.298} & 0.112 & 0.191\\ & Calibration & 1.288 & 1.479 & 1.643 & \textbf{0.84} & - & 2.538\\
\midrule
\multirow{3}{*}{malonaldehyde} & Energy & 1.493 & 0.942 & 0.130 & \textbf{0.106} & 0.111 & 0.098\\ & Forces & 1.467 & 1.185 & 0.535 & \textbf{0.465} & 0.461 & 0.368\\ & Calibration & 1.257 & 1.319 & 1.580 & \textbf{1.208} & - & 2.401\\
\midrule
\multirow{3}{*}{napthalene} & Energy & 2.739 & 1.645 & 0.346 & \textbf{0.132} & 0.131 & 0.124\\ & Forces & 1.385 & 1.119 & 0.357 & \textbf{0.263} & 0.255 & 0.211\\ & Calibration & 1.280 & 1.330 & \textbf{1.258} & 1.297 & - & 2.601\\
\midrule
\multirow{3}{*}{salicylic acid} & Energy & 2.301 & 1.310 & 0.165 & \textbf{0.132} & 0.141 & 0.132\\ & Forces & 1.853 & 1.535 & 0.607 & \textbf{0.457} & 0.465 & 0.378\\ & Calibration & 1.266 & 1.389 & 1.527 & \textbf{0.801} & - & 2.519\\
\midrule
\multirow{3}{*}{toluene} & Energy & 1.662 & 0.621 & 0.202 & \textbf{0.106} & 0.108 & 0.099\\ & Forces & 1.404 & 1.155 & 0.402 & \textbf{0.282} & 0.296 & 0.230\\ & Calibration & 1.269 & 1.535 & 1.472 & \textbf{1.042} & - & 2.649\\
\midrule
\multirow{3}{*}{uracil} & Energy & 1.759 & 0.959 & 0.168 & \textbf{0.123} & 0.119 & 0.115\\ & Forces & 1.945 & 1.656 & 0.480 & \textbf{0.341} & 0.326 & 0.265\\ & Calibration & \textbf{1.262} & 1.396 & 1.524 & 1.370 & - & 2.725\\
\bottomrule
\end{tabular}

%% file: tables/qm7x/qm7x_dimenet.tex
\centering
\begin{tabular}{lccccc}
\toprule
{} & MC-Dropout 20\% & MC-Dropout 1\% & Evidential & \oursacro{} & \oursacro{} + Drop 5\%\\
\midrule
Energy ID  &   0.240 & 0.090 & \textbf{0.020} & 0.049 & 0.0209\\
Forces ID  &   0.009 & \textbf{0.004} & 0.023 & 0.047 & 0.0045 \\
Energy OOD &   5.242 & 5.277 & 5.172 & 5.530 & 5.230\\
Forces OOD &   1.266 & 1.282 & 1.367 & 1.361 & 1.293 \\
\midrule
Calibration    &   \textbf{1.279} & 1.285 & 1.629 & 2.596 & 2.596 \\
\midrule
AUC-ROC energy &   0.511 & 0.504 & 0.587 & \textbf{0.795} & 0.7913\\
AUC-ROC forces (trace)  & \textbf{0.905} & 0.858 & \noindent\rule{0.8cm}{0.6pt} & 0.729 & 0.8278\\
\bottomrule
\end{tabular}

%% file: tables/qm7x/qm7x_dimenet_69.tex
\centering
\begin{tabular}{lccccc}
\toprule
{} & MC-Dropout 20\% & MC-Dropout 1\% & Evidential & \oursacro{} & \oursacro{} + Drop 1\%\\
\midrule
Energy ID  &   0.165 & 0.077 & \textbf{0.020} & 0.049 & 0.0265\\
Forces ID  &   0.007 & \textbf{0.004} & 0.023 & 0.047 & 0.0061 \\
Energy OOD &   5.022 & 5.063 & 5.172 & 5.53 & 5.23\\
Forces OOD &   1.27 & 1.271 & 1.367 & 1.361 & 1.293 \\
\midrule
Calibration    &   1.367 & \textbf{1.296} & 1.629 & 2.596 & 2.596 \\
\midrule
AUC-ROC energy &   0.513 & 0.521 & 0.587 & \textbf{0.795} & 0.74\\
AUC-ROC forces (trace)  & \textbf{0.911} & 0.865 & \noindent\rule{0.8cm}{0.6pt} & 0.729 & 0.854 \\
\bottomrule
\end{tabular}

%% file: tables/qm7x/qm7x_other_models.tex
\begin{tabular}{lcccc}
\toprule
 & \multicolumn{2}{c}{MC-Dropout} & \multicolumn{2}{c}{Evidential}  \\
\midrule
             & SchNet & NequIP  & SchNet & NequIP \\
 \midrule
Energy ID    & 0.44 & 0.158  & 0.027 & 0.019 \\
Forces ID    & 0.025 & 0.019  & 0.027 & 0.009 \\
Energy OOD   & 5.292 & $37.751^{*}$  & 5.322 & 21.24 \\
Forces OOD   & 1.271 & $26.139^{*}$  & 1.319 & 84.441 \\
\midrule
Calibration  & 1.346 & 1.272  & 1.670 & 1.549 \\
 \midrule
AUC-ROC energy    & 0.501 & 0.627  & 0.741 & 0.746\\
AUC-ROC forces (trace)    & 0.8 & 0.988  & \noindent\rule{0.8cm}{0.6pt} & \noindent\rule{0.8cm}{0.6pt}\\
\bottomrule
\end{tabular}

%% file: tables/qm7x/qm7x_other_models_69samples.tex
\begin{tabular}{lcccc}
\toprule
 & \multicolumn{2}{c}{MC-Dropout} & \multicolumn{2}{c}{Evidential}  \\
\midrule
 & SchNet & NequIP  & SchNet & NequIP \\
 \midrule
Energy ID    & 0.423 & 0.109  & 0.027 & 0.019 \\
Forces ID    & 0.023 & 0.017  & 0.027 & 0.009 \\
Energy OOD   & 5.291 & $36.253^{*}$  & 5.322 & 21.24 \\
Forces OOD  & 1.271 & $25.937^{*}$  & 1.319 & 84.441 \\
\midrule
Calibration  & 1.274 & 1.329  & 1.670 & 1.549 \\
 \midrule
AUC-ROC energy    & 0.478 & 0.68  & 0.741 & 0.746\\
AUC-ROC forces (trace)    & 0.81 & 0.992  & \noindent\rule{0.8cm}{0.6pt} & \noindent\rule{0.8cm}{0.6pt}\\
\bottomrule
\end{tabular}

%% file: tables/qm7x/dropout_exp_dimenet.tex
\begin{tabular}{lccccc}
\toprule
{Dropout Rate}            & 1\% & 5\% & 10\% & 15\% & 20\%  \\
\midrule
Energy ID             & 0.09 & 0.134 & 0.259 & 0.206 & 0.24  \\
Forces ID             & 0.004 & 0.005 & 0.012 & 0.008 & 0.009  \\
Energy OOD            & 5.277 & 5.319 & 4.978 & 5.245 & 5.242  \\
Forces OOD             & 1.282 & 1.288 & 1.675 & 1.273 & 1.266  \\
\midrule
Calibration           &  1.285 & 1.275 & 1.273 & 1.271 & 1.279  \\
\midrule
AUC-ROC energy         &  0.504 & 0.511 & 0.55 & 0.517 & 0.511 \\
AUC-ROC forces (trace) &  0.858 & 0.872 & 0.886 & 0.904 & 0.905  \\
\bottomrule
\end{tabular}

%% file: tables/qm7x/dropout_exp_schnet.tex
\begin{tabular}{lccccc}
\toprule
{Dropout Rate}                & 1\% & 5\% & 10\% & 15\% & 20\%  \\
\midrule
Energy ID                 & 0.177 & 0.256 & 0.359 & 0.430 & 0.440  \\
Forces ID                 & 0.012 & 0.015 & 0.019 & 0.022 & 0.025  \\
Energy OOD                & 5.126 & 5.135 & 5.036 & 5.235 & 5.292  \\
Forces OOD               & 1.255 & 1.262 & 1.260 & 1.260 & 1.271  \\
\midrule
Calibration              &  1.318 & 1.328 & 1.347 & 1.332 & 1.346  \\
\midrule
AUC-ROC energy           &  0.492 & 0.493 & 0.496 & 0.484 & 0.501 \\
AUC-ROC forces (trace)   &  0.816 & 0.845 & 0.826 & 0.811 & 0.800 \\
\bottomrule
\end{tabular}